\documentclass[fleqn,usenatbib]{mnras}

\usepackage{newtxtext,newtxmath}
\usepackage[T1]{fontenc}

\DeclareRobustCommand{\VAN}[3]{#2}
\let\VANthebibliography\thebibliography
\def\thebibliography{\DeclareRobustCommand{\VAN}[3]{##3}\VANthebibliography}

\usepackage{graphicx}	% Including figure files
\usepackage{amsmath}	% Advanced maths commands
\usepackage{orcidlink}	% Clickable ORCID iDs beside author names

\usepackage{cases}

\usepackage{caption}

\usepackage{threeparttable}
\usepackage{longtable,booktabs}
\usepackage{needspace}

\usepackage{multirow}

\usepackage{longtable}

\title[Bulge kinematics by HST proper motions]{Kinematic Map of the Galactic Bulge Using HST Proper Motions}

\author[Zhenhao Wang et al.]{%
Zhenhao Wang\,\orcidlink{0009-0000-0099-5325},$^{1}$%\thanks{E-mail: wangzhenhao@westlake.edu.cn}
Shude Mao\,\orcidlink{0000-0001-8317-2788},$^{1}$\thanks{E-mail: shude.mao@westlake.edu.cn}
Hongjing Yang\,\orcidlink{0000-0003-0626-8465}$^{1}$,
Andrzej Udalski\,\orcidlink{0000-0001-5207-5619}$^{2}$,
Weicheng Zang\,\orcidlink{0000-0001-6000-3463}$^{1}$
\newauthor
and T. Dex Bhadra\,\orcidlink{0009-0002-6097-9030}$^{3}$
\\
$^{1}$Department of Astronomy, Westlake University, Hangzhou 310030, Zhejiang Province, People's Republic of China\\
$^{2}$Astronomical Observatory of the University of Warsaw, Al. Ujazdowskie 4, PL-00-478 Warszawa, Poland\\
$^{3}$Department of Astronomy and Astrophysics, University of Toronto,  Toronto, ON M5S 3H4, Canada
}

\date{Accepted XXX. Received YYY; in original form ZZZ}

\pubyear{2026}

\begin{document}
\label{firstpage}
\pagerange{\pageref{firstpage}--\pageref{lastpage}}
\maketitle

\begin{abstract}
  Constraining the Galactic bulge's non-axisymmetric dynamics requires proper motions in deep fields across many sightlines, yet crowding and extinction limit such measurements. Using multi-epoch observations of 27 fields in the southern Galactic bulge (\(-6^\circ < l < 4^\circ\), \(-5^\circ < b < -1^\circ\)) from the ultraviolet–visible (UVIS) channel of Wide Field Camera 3 (WFC3) on the Hubble Space Telescope (HST) with time baselines of 3 to 13 years, and combining them with the absolute proper motion reference frame provided by \textit{Gaia} Data Release 3 (DR3), we measure absolute proper motions for \(\sim400,000\) stars. Colour-magnitude diagrams are used to separate bulge stars from foreground disc stars, identify red clump stars and correct for interstellar extinction and reddening. Statistical analysis shows that the mean proper motion in Galactic latitude is consistent with that of the central supermassive black hole Sgr A*, while the mean proper motion in Galactic longitude is affected by disc stars but agrees with Sgr A* after extrapolation to the Galactic plane. Proper motion dispersions increase towards the Galactic centre and are asymmetric across Galactic longitude, with a steeper gradient for \(l<0^\circ\), revealing the bulge's non-axisymmetric bar structure. The velocity distribution shows clear anisotropy (\(\sigma_l/\sigma_b=1.131\pm0.005\)). The normalised covariance is negative in the \(l>0^\circ\) region (\(\sim -0.085\)) and close to zero in the \(l<0^\circ\) region. The proper motion distributions are non-Gaussian in both Galactic longitude and latitude, mainly due to contamination by foreground disc stars.
\end{abstract}


%%%%%%%%%%%%%%%%%%%%%%%%%%%%%%%%%%%%%%%%%%%%%%%%%%

%%%%%%%%%%%%%%%%% BODY OF PAPER %%%%%%%%%%%%%%%%%%

% !TEX root = ../mnras_template.tex

\section{Introduction}\label{chap:01introduction}

The Milky Way is the only spiral galaxy that can be resolved into individual stars, providing not only the number density but also the six-dimensional phase-space information (three-dimensional positions and velocities) and intrinsic properties such as chemical abundances and ages for vast numbers of stars. This makes it an irreplaceable near-field laboratory for tracing galaxy formation and evolution. The Galactic bulge, as the densest stellar assembly in the central region, preserves imprints of both early formation and secular evolution, and its structure and kinematics are directly linked to the formation history of the bar, the boxy/peanut-shaped bulge, and a possible classical bulge component. See \citet{BlandHawthorn2016GalaxyContext} for a review.

Proper motions provide the two-dimensional transverse velocities on the plane of the sky, complementing the line-of-sight radial velocities to more completely constrain the velocity ellipsoid, rotation, dispersions, and their spatial gradients. For bulge stars at several kiloparsecs, milliarcsecond-per-year precision suffices to resolve internal motions. Mean proper motions reflect solar motion and the mixing of populations along different lines of sight; proper motion dispersions characterise random motions; and the dispersion ratio and covariance reveal anisotropy, tilt of the velocity ellipsoid, and possible non-axisymmetric structures. These quantities are critical for constraining the dynamics of the Galactic bar, disc contamination, and the composition of different bulge populations.

In studies of the formation mechanism of the Galactic bulge, two main classes of theories currently exist: classical bulge formation and pseudo-bulge formation driven by disc instability. Classical bulges are thought to form through rapid gravitational collapse and mergers with massive clumps or satellite galaxies in the early Universe \citep{Eggen1962GalaxyCollapse,Noguchi1999BulgeFormationClumps,Aguerri2001BulgeGrowthMergers}, and are morphologically pressure-supported, spherical or spheroidal structures, with populations characterised by old ages (>10 Gyr), low metallicities, and high velocity dispersions \citep{Barbuy2018BulgeChemodynamics}. Pseudo-bulges, on the other hand, arise from the secular evolution of the galactic disc itself—the inner disc forms a large-scale bar through bar instability, and subsequently the bar thickens via vertical buckling instability, eventually shaping a boxy/peanut-shaped (B/P) bulge \citep{Combes1981BarFormation, Raha1991BarInstability, Athanassoula2005BoxPeanutBulges}. Kinematically, such bulges are characterised by near-cylindrical rotation and relatively low velocity dispersions, with stellar ages spanning a wide range and metallicity distribution peaks often near solar values \citep{KK2004SecularPseudobulges}.

A wealth of observations indicates that the Galactic bulge overall exhibits a prominent boxy/peanut shape. The distributions of red clump (RC) stars clearly show an X-shaped structure \citep{McWilliam2010RedClumpsXshapedBulge, Nataf2010SplitRedClump, Saito2011MappingXshapedBulge, Wegg2013Bulge3D}. Using RC star counts from Optical Gravitational Lensing Experiment III \citep[OGLE-III;][]{Szymanski2011OGLEIII} data, \citet{Cao2013PhotometricBar} fitted a triaxial photometric model that constrained the bar's shape and orientation. These morphologies support a pseudo-bulge formed via disc secular evolution. However, growing evidence indicates the presence of a small number of old, kinematically hot populations in the innermost region and at the metal-poor end, whose chemical abundance patterns are closer to those of spheroidal components, implying that a low-mass classical bulge component may be nested inside the pseudo-bulge \citep{Babusiaux2010InsightsBulgeFormation, Zoccali2014GIBSKinematics}. Therefore, the current picture favours a Galactic bulge that is predominantly a pseudo-bulge, with a possible minor classical bulge at the centre.

Early kinematic constraints on the bulge mainly came from radial-velocity surveys. The Bulge Radial Velocity Assay \citep[BRAVA;][]{Rich2007BRAVARotationCurve,Howard2008BRAVARotationCurve,Kunder2012BRAVA} observed about 10,000 M giants and established that the bulge exhibits near-cylindrical rotation, a key signature of a bar-driven pseudo-bulge rather than a merger-built classical bulge \citep{Shen2010PureDisk}, together with trends of velocity dispersion with position \citep{Howard2009CylindricalRotation}. Combining BRAVA kinematics with an existing $N$-body model, \citet{Long2013MadeToMeasure} adapted made-to-measure modelling to the rotating Galactic bar. The Abundances and Radial velocity Galactic Origins Survey \citep[ARGOS;][]{Freeman2013ARGOS} extended the sample to about 28,000 RC stars with metallicity information; \citet{Ness2013ARGOSBulgeKinematics} found that the dominant metal-rich component shows strong cylindrical rotation and bar-like kinematics, while the metal-poor tail is hotter and more slowly rotating, more consistent with the inner halo or a classical bulge. The Apache Point Observatory Galactic Evolution Experiment \citep[APOGEE;][]{Majewski2017APOGEE} obtained infrared spectra for hundreds of thousands of bulge stars, extending the chemical–kinematic picture into regions heavily obscured by dust that are difficult to probe optically \citep{Queiroz2021APOGEEBarBulge}. These surveys provide important constraints on the bulge's dynamical state, but lacking transverse velocities, they still cannot fully break the degeneracies among formation models.

In nearby, data-rich regions of the Galactic disc, high-precision proper motions combined with radial velocities have already demonstrated strong diagnostic power: \citet{Zari2025DiskKinematics} measured large-scale alternating radial flows and a "kinematic fractionation" in young OB stars, while \citet{Palicio2023SpiralFeatures} used radial actions of tens of millions of stars to link arc-like structures to spiral arms and the bar's outer Lindblad resonance. Using classical Cepheids, \citet{Mroz2019RotationCurveMilkyWay} were able to trace the Milky Way rotation curve across the disc by \textit{Gaia} \citep{GaiaCollaboration2016gaia,GaiaCollaboration2023gaia} and OGLE proper motion data.

Towards the bulge, however, proper motion measurements face strong, rapidly varying extinction and extreme crowding \citep{Nataf2013ReddeningExtinction}. \citet{Spaenhauer1992BulgePMs} obtained the first bulge proper motion data of bulge stars in Baade's Window using Palomar telescope. After that, \citet{Sumi2004OGLEII} compiled proper motions for 5,080,236 stars in 49 OGLE-II \citep{Udalski2000OGLEII} bulge fields, from which \citet{Rattenbury2007BulgePMs} measured proper motion dispersions of 577,736 red clump giants and found the observed velocity ellipsoid to be less anisotropic than predicted by their particle models. In the near-infrared, the VISTA Variables in the Via Lactea survey \citep[VVV;][]{Minniti2010VVV} and its VVV Infrared Astrometric Catalogue \citep[VIRAC;][]{Smith2018VIRAC} provide relative proper motions for about $10^8$ sources; combined with \textit{Gaia}, \citet{Clarke2019VIRAC3D} built a three-dimensional view of bar/bulge kinematics in proper motion and \citet{Clarke2022PatternSpeed} measured the bar pattern speed. \citet{Steven2022BulgePMRotationCurves} ran an N-body hydrodynamics simulation and found the proper motion rotation curves of young bulge stars showing a steeper slope relative to old bulge stars, providing predictions for future proper motion observations with the Hubble Space Telescope (HST) and the Legacy Survey of Space and Time \citep[LSST;][]{Ivezic2019LSST}.

Among the proper motion studies of bulge area, HST holds a significant position. The stable point spread function (PSF) and 0.04~arcsec pixel$^{-1}$ scale of HST make it particularly suited to precise photometry and astrometry in crowded fields. In the beginning of this century, \citet{Kuijken2002BulgePMs} used Wide Field and Planetary Camera 2 (WFPC2) on HST to measure proper motion dispersions in two bulge fields, and \citet{Kozlowski2006kinematics} presented the first large-area HST proper motion survey in the bulge, measuring relative proper motions of 15,863 stars in 35 fields and, combining with \citet{Kuijken2002BulgePMs}, detecting spatial gradients of the anisotropy ratio and the transverse covariance term $\sigma_{lb}/(\sigma_l\sigma_b)\approx-0.10$. Using deep field data from Sagittarius Window Eclipsing Extrasolar Planet Search (SWEEPS), \citet{Clarkson2008BulgeProperMotions} reached $\sim0.1$~mas yr$^{-1}$ for about 15,000 stars and kinematically separated bulge and disc populations; \citet{Clarkson2011BulgeBlueStragglers} further constrained the disc-to-bulge mass ratio and bar-orbit fraction. Using observations from the Advanced Camera for Surveys (ACS) on HST with 8--9-yr baselines, \citet{Soto2023HSTProperMotionsGalacticBar} measured about 100,000 proper motions in four far-side (negative-longitude) fields, reaching $<0.5$~mas yr$^{-1}$ for $m_{\text{F814W}}<23$ and revealing significant bulge rotation that extends to $l\approx-8^{\circ}$. \citet{Luna2023AstrometryBulge} assessed the reliability of proper motions in crowded bulge fields, showing that VIRAC version 2 \citep[VIRAC2;][]{Smith2025VIRAC2} agrees with HST and complements \textit{Gaia} Data Release 3 (DR3), whose uncertainties are underestimated by up to a factor of $\sim4$ in the densest fields. Near the Galactic centre, \citet{Libralato2021GC2DKinematics} achieved better than $0.5$~mas yr$^{-1}$ and identified candidate hypervelocity stars, demonstrating the value of HST archival data in dense fields. More recently, \citet{Terry2026HSTBulgeWideSurvey} presented a wide-field HST imaging survey of the Galactic bulge that overlaps the Galactic Bulge Time-Domain Survey
(GBTDS) planned for the Nancy Grace Roman Space Telescope
\citep[Roman;][]{RomanROTAC2025} and provides a resource for future studies of bulge stellar populations and dynamics.

Previous HST proper motion studies towards bulge area were mainly limited to a handful of fields, which cannot reveal trends in large scale. Other studies with more fields were limited by small stellar samples due to earlier instruments, high uncertainties due to short time baselines, and relative-only calibration. While this study can now overcome these by the
ultraviolet–visible (UVIS) channel of Wide Field Camera 3 (WFC3), decade-long baselines, and absolute reference frame from sky survey missions such as \textit{Gaia} and OGLE. From the processing side, charge transfer efficiency (CTE) corrected \texttt{\_flc} data and the \texttt{hst1pass} software \citep{Anderson2010EmpiricalCTE,Anderson2021wfcuvisCTE,Anderson2022hst1pass} provide mature corrections for geometric distortion and PSF effects. Therefore, we can now measure absolute proper motions for a larger number of stars covering a larger bulge region with high precision and accuracy, and study their kinematics in detail.

The structure of this paper is as follows: Section~\ref{chap:02data} describes the HST field selection, data processing pipeline, proper motion measurement methods, and error estimation; Section~\ref{chap:03statistic_method} explains the statistical definitions of mean proper motion, dispersion, anisotropy and covariance, and introduces bootstrap error estimation and colour–magnitude diagram (CMD) analysis; Section~\ref{chap:04results} presents the statistical results of proper motions in different bulge regions, and discusses their spatial distributions, comparisons with previous studies and possible physical implications; Section~\ref{chap:05conclusion} summarises the main work and conclusions of this paper.
% !TEX root = ../mnras_template.tex

\begin{figure*}
  \centering
  \includegraphics[width=1.0\linewidth]{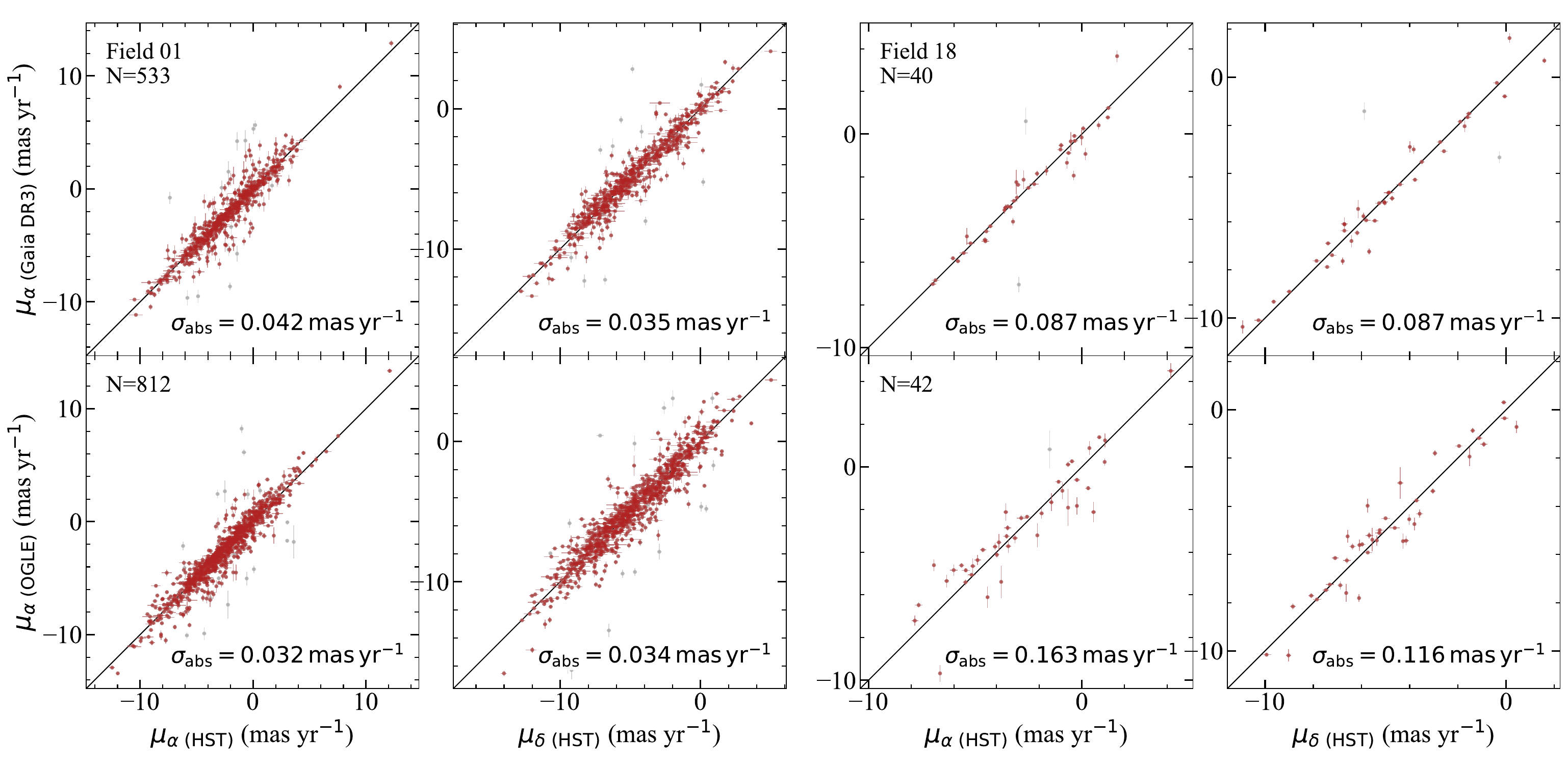}
  \caption{Proper motion comparison between HST and \textit{Gaia} DR3 (top) and OGLE/Uranus (bottom) after absolute proper motion calibration. The proper motions are in right ascension ($\alpha$) and declination ($\delta$). $\sigma_{\text{abs}}$ is the standard error of absolute proper motion calibration calculated by equation~\eqref{eq:abs_calib_error}. N stands for the number of stars eventually used for the calibration (red), while the grey points are excluded due to the $4\sigma_{\Delta\mu}$ clipping. The black lines indicate the $1:1$ relation. The left 2 panels show data from field 01 ($l=1.256^\circ$, $b=-2.651^\circ$), and the right 2 panels show data from field 18 ($l=1.277^\circ$, $b=-2.551^\circ$). Some grey points close to the $1:1$ line are due to the $4\sigma_{\Delta\mu}$ clipping in the other direction.}
  \label{fig:pm_gaia_ogle}
\end{figure*}

\section{Fields and Data Reduction}\label{chap:02data}

\subsection{HST fields}

We query the Mikulski Archive for Space Telescopes (MAST) for HST observations in the Galactic bulge region ($-7^\circ < l < 7^\circ$, $-7^\circ < b < 0^\circ$) taken with the WFC3/UVIS camera. In this region, F814W and F606W are the two most frequently used filters, having accumulated a large amount of highly consistent archival data. Moreover, these two bands have well-established PSF models that support high-precision astrometry, and they are also the standard filters in the instrument stability and contamination monitoring programmes \citep{Gosmeyer2014UpdateOT}. Therefore, this study is based on observations in these two bands.

We screen all F814W observations with the following criteria: multiple observations in the same field separated by more than 3 years; if a field has more than two observations, we select the two with the longest time baseline. Finally, we identify 27 qualifying fields, with time baselines of $3\sim13$ years, covering the range $-6^\circ < l < 4^\circ$, $-5^\circ < b < -1^\circ$, essentially covering the entire southern Galactic bulge. The WFC3/UVIS camera consists of two Charge-Coupled Devices (CCDs) forming a total field of view of about $162\times162$~arcsec$^{2}$. But in practice, to reduce data volume and readout time, smaller aperture sizes are often used, so the actual field size may be smaller than the full-frame field. The field sizes in the selected observations range from $41\times41$~arcsec$^{2}$ to $162\times162$~arcsec$^{2}$. At the same time, the roll angle and the overlapping area between the two epochs vary from field to field, so the effective coverage differs among fields.

Table~\ref{tab:fields_summary} summarises the basic information of the 27 fields in this study, including right ascension (RA), declination (Dec), galactic longitude ($l$), galactic latitude ($b$), filters, observation IDs, observation years, and exposure times. For each field, the two selected observations are designated as Epoch 1 (the earlier observation) and Epoch 2 (the later observation). Among the 27 fields, 16 have F606W observations at Epoch 1. In this study, only F814W data are used for astrometry and proper motion measurements; F606W is used solely to obtain stellar colour information, to assist in stellar classification and in constructing colour–magnitude diagrams (CMDs). Therefore, for F606W observations, only the exposure with the longest exposure time is analysed.

\begin{table*}
  \centering
  \caption{HST fields information. Right ascension (RA) and declination (Dec) are in the International Celestial Reference System (ICRS). RA, Dec, galactic longitude ($l$), and galactic latitude ($b$) are all given in degrees.}\label{tab:fields_summary}
  \begin{tabular}{ccccccp{2.7cm}}
    \toprule
    Field ID & RA, Dec (ICRS) & $l$, $b$ & Filters & Observation ID & Year & Exposure time (s) \\
    \midrule
    % Field 01 (no F606W)
    \multirow{2}{*}{01} & \multirow{2}{*}{269.7466, -29.2049} & \multirow{2}{*}{1.2558, -2.6506} & F814W & ib3m03010 & 2010.36 & $348\times4$, $740$ \\
                         & & & F814W & icp011teq & 2015.58 & $165$ \\
    \midrule
    % Field 02 (has F606W)
    \multirow{3}{*}{02} & \multirow{3}{*}{269.8558, -27.3587} & \multirow{3}{*}{2.9068, -1.8161} & F814W & ibcw04010 & 2010.56 & $375\times4$ \\
                         & & & F606W & ibcw04a6q & 2010.56 & $390$ \\
                         & & & F814W & iepz20010 & 2022.43 & $300\times2$ \\
    \midrule
    % Field 03 (has F606W)
    \multirow{3}{*}{03} & \multirow{3}{*}{269.6116, -26.8401} & \multirow{3}{*}{3.2483, -1.3693} & F814W & ibcw05zwq & 2009.75 & $20$ \\
                         & & & F606W & ibcw05a3q & 2009.75 & $100$ \\
                         & & & F814W & ic7m01010 & 2013.46 & $275\times4$ \\
    \midrule
    % Field 04 (has F606W)
    \multirow{3}{*}{04} & \multirow{3}{*}{271.2359, -29.0384} & \multirow{3}{*}{2.0420, -3.7015} & F814W & ibcw10010 & 2010.43 & $65\times6$, $15$ \\
                         & & & F606W & ibcw10c7q & 2010.43 & $60$ \\
                         & & & F814W & iepz19010 & 2022.29 & $65\times2$ \\
    \midrule
    % Field 05 (has F606W)
    \multirow{3}{*}{05} & \multirow{3}{*}{267.9175, -29.8906} & \multirow{3}{*}{-0.1387, -1.6210} & F814W & ibnc06020 & 2011.60 & $75\times3$, $60$, $120$ \\
                         & & & F606W & ibnc06f6q & 2011.60 & $75$ \\
                         & & & F814W & ieos02010 & 2022.70 & $307\times6$ \\
    \midrule
    % Field 06 (has F606W)
    \multirow{3}{*}{06} & \multirow{3}{*}{268.9826, -30.5610} & \multirow{3}{*}{-0.2515, -2.7540} & F814W & ibnc07020 & 2011.62 & $20\times3$, $40\times4$ \\
                         & & & F606W & ibnc07ahq & 2011.62 & $30$ \\
                         & & & F814W & iepz13010 & 2022.29 & $60\times2$ \\
    \midrule
    % Field 07 (has F606W)
    \multirow{3}{*}{07} & \multirow{3}{*}{267.8558, -30.4097} & \multirow{3}{*}{-0.6126, -1.8398} & F814W & ibnc08020 & 2011.72 & $75\times5$ \\
                         & & & F606W & ibnc08ojq & 2011.72 & $75$ \\
                         & & & F814W & iepz14010 & 2022.11 & $250\times2$ \\
    \midrule
    % Field 08 (no F606W)
    \multirow{2}{*}{08} & \multirow{2}{*}{271.5222, -30.7326} & \multirow{2}{*}{0.6769, -4.7401} & F814W & ibqt01010 & 2011.80 & $85\times4$ \\
                         & & & F814W & iepz01010 & 2022.31 & $340\times2$ \\
    \midrule
    % Field 09 (no F606W)
    \multirow{2}{*}{09} & \multirow{2}{*}{269.1082, -32.2374} & \multirow{2}{*}{-1.6512, -3.6861} & F814W & ibqt02010 & 2012.13 & $105\times6$ \\
                         & & & F814W & iepz02010 & 2022.21 & $215\times2$ \\
    \midrule
    % Field 10 (no F606W)
    \multirow{2}{*}{10} & \multirow{2}{*}{268.5605, -34.7781} & \multirow{2}{*}{-4.0855, -4.5626} & F814W & ibqt03010 & 2012.14 & $70\times8$ \\
                         & & & F814W & iepz33010 & 2022.14 & $15\times2$ \\
    \midrule
    % Field 11 (no F606W)
    \multirow{2}{*}{11} & \multirow{2}{*}{272.0158, -27.1501} & \multirow{2}{*}{4.0309, -3.3877} & F814W & ibqt04010 & 2012.14 & $205\times8$ \\
                         & & & F814W & iepz32010 & 2023.60 & $300\times2$ \\
    \midrule
    % Field 12 (has F606W)
    \multirow{3}{*}{12} & \multirow{3}{*}{269.8336, -29.2726} & \multirow{3}{*}{1.2348, -2.7501} & F814W & ibts01040 & 2011.80 & $410\times2$ \\
                         & & & F606W & ibts01y7q & 2011.80 & $410$ \\
                         & & & F814W & ic8x61030 & 2014.81 & $450$, $350$ \\
    \midrule
    % Field 13 (has F606W)
    \multirow{3}{*}{13} & \multirow{3}{*}{269.9121, -29.2645} & \multirow{3}{*}{1.2758, -2.8056} & F814W & ibts02040 & 2011.80 & $410\times2$ \\
                         & & & F606W & ibts02ylq & 2011.80 & $410$ \\
                         & & & F814W & iepz36010 & 2022.43 & $300\times2$ \\
    \midrule
    % Field 14 (has F606W)
    \multirow{3}{*}{14} & \multirow{3}{*}{269.8429, -29.3282} & \multirow{3}{*}{1.1905, -2.7848} & F814W & ibts03040 & 2011.80 & $410\times2$ \\
                         & & & F606W & ibts03yzq & 2011.80 & $410$ \\
                         & & & F814W & ic8x63030 & 2014.81 & $450$, $350$ \\
    \midrule
    % Field 15 (has F606W)
    \multirow{3}{*}{15} & \multirow{3}{*}{269.7787, -29.3306} & \multirow{3}{*}{1.1606, -2.7373} & F814W & ibts04040 & 2011.80 & $410\times2$ \\
                         & & & F606W & ibts04aaq & 2011.80 & $410$ \\
                         & & & F814W & ic8x64030 & 2014.81 & $450$, $350$ \\
    \midrule
    % Field 16 (has F606W)
    \multirow{3}{*}{16} & \multirow{3}{*}{269.6715, -29.1235} & \multirow{3}{*}{1.2939, -2.5532} & F814W & ibts05040 & 2011.81 & $410\times2$ \\
                         & & & F606W & ibts05k2q & 2011.81 & $410$ \\
                         & & & F814W & iepz25010 & 2022.13 & $130\times2$ \\
    \midrule
    % Field 17 (has F606W)
    \multirow{3}{*}{17} & \multirow{3}{*}{269.6891, -29.2012} & \multirow{3}{*}{1.2341, -2.6052} & F814W & ibts07040 & 2011.81 & $410\times2$ \\
                         & & & F606W & ibts07lsq & 2011.81 & $410$ \\
                         & & & F814W & iepz29010 & 2022.27 & $120\times2$ \\
    \midrule
    % Field 18 (has F606W)
    \multirow{3}{*}{18} & \multirow{3}{*}{269.6593, -29.1367} & \multirow{3}{*}{1.2771, -2.5505} & F814W & ibts09040 & 2012.20 & $410\times2$ \\
                         & & & F606W & ibts09xsq & 2012.20 & $410$ \\
                         & & & F814W & iffp01020 & 2025.31 & $69\times16$ \\
    \midrule
    % Field 19 (has F606W)
    \multirow{3}{*}{19} & \multirow{3}{*}{269.9019, -29.2965} & \multirow{3}{*}{1.2436, -2.8137} & F814W & ibts34020 & 2012.50 & $400\times2$ \\
                         & & & F606W & ibts34pgq & 2012.50 & $400$ \\
                         & & & F814W & iepz26010 & 2022.39 & $340\times2$ \\
    \midrule
    % Field 20 (no F606W)
    \multirow{2}{*}{20} & \multirow{2}{*}{269.7060, -30.1969} & \multirow{2}{*}{0.3763, -3.1129} & F814W & ic9i02010 & 2013.77 & $70\times8$ \\
                         & & & F814W & iepz34010 & 2022.27 & $10\times2$ \\
    \bottomrule
  \end{tabular}
\end{table*}

% 第二个 table*：Field 19–27
\begin{table*}
  \ContinuedFloat
  \centering
  \caption{HST fields information (continued).}
  \begin{tabular}{ccccccp{2.7cm}}
    \toprule
    Field ID & RA, Dec (ICRS) & $l$, $b$ & Filters & Observation ID & Year & Exposure time (s) \\
    \midrule
    % Field 21 (no F606W)
    \multirow{2}{*}{21} & \multirow{2}{*}{269.7451, -29.2288} & \multirow{2}{*}{1.2345, -2.6613} & F814W & icp009010 & 2015.57 & $165\times2$ \\
                         & & & F814W & iepz23010 & 2023.33 & $340\times2$ \\
    \midrule
    % Field 22 (no F606W)
    \multirow{2}{*}{22} & \multirow{2}{*}{265.3997, -34.5554} & \multirow{2}{*}{-5.2341, -2.2188} & F814W & idju01010 & 2017.64 & $150\times4$, $250\times4$ \\
                         & & & F814W & ieoj01010 & 2023.21 & $318\times7$ \\
    \midrule
    % Field 23 (no F606W)
    \multirow{2}{*}{23} & \multirow{2}{*}{268.5398, -28.7479} & \multirow{2}{*}{1.1217, -1.5081} & F814W & idju04020 & 2017.81 & $30\times3$, $50\times10$, $60\times3$ \\
                         & & & F814W & ied253010 & 2021.58 & $125\times9$ \\
    \midrule
    % Field 24 (no F606W)
    \multirow{2}{*}{24} & \multirow{2}{*}{269.4963, -30.7152} & \multirow{2}{*}{-0.1640, -3.2138} & F814W & idsa02010 & 2018.40 & $84\times15$ \\
                         & & & F814W & iepz11010 & 2023.34 & $300\times2$ \\
    \midrule
    % Field 25 (no F606W)
    \multirow{2}{*}{25} & \multirow{2}{*}{272.0193, -29.7316} & \multirow{2}{*}{1.7647, -4.6341} & F814W & idsa03020 & 2018.39 & $62\times6$ \\
                         & & & F814W & iepz22010 & 2022.17 & $150\times2$ \\
    \midrule
    % Field 26 (has F606W)
    \multirow{3}{*}{26} & \multirow{3}{*}{272.5186, -27.8671} & \multirow{3}{*}{3.6159, -4.1235} & F814W & idzi07020 & 2019.58 & $175\times9$ \\
                         & & & F606W & idzi07shq & 2019.58 & $100$ \\
                         & & & F814W & ied206010 & 2023.31 & $212\times3$ \\
    \midrule
    % Field 27 (has F606W)
    \multirow{3}{*}{27} & \multirow{3}{*}{271.4794, -30.3369} & \multirow{3}{*}{1.0066, -4.5166} & F814W & ie8j01010 & 2020.20 & $87\times16$ \\
                         & & & F606W & ie8j01y8q & 2020.20 & $89$ \\
                         & & & F814W & iffe02020 & 2025.43 & $80\times14$ \\
    \bottomrule
  \end{tabular}
\end{table*}

\subsection{Data reduction and proper motion measurement}

The data product selected for this study is the \texttt{\_flc} files, which are the products of the standard WFC3/UVIS calibration pipeline. They have undergone basic calibration steps including bias correction, dark-current correction, and flat-field correction. Compared with the same-level \texttt{\_flt} products, the \texttt{\_flc} data include an additional charge transfer efficiency (CTE) correction. The CTE loss problem refers to the fact that during the CCD readout process, charge may be delayed when being transferred from pixels to the readout register. This is especially problematic for space telescopes such as HST, whose CCD pixels are exposed to a high-energy radiation environment for long periods, making them more susceptible to CTE loss, which in turn causes trailing and blurring of stellar images and thus affects astrometric and photometric accuracy \citep{Ubeda2012ACSWFCCTE}. CTE loss has been found in several HST instruments. To address this issue for the ACS Wide Field Channel (WFC), \citet{Anderson2010EmpiricalCTE} proposed an empirical method, which was later adapted to the WFC3/UVIS camera for CTE correction and incorporated into the standard calibration pipeline \citep{Anderson2021wfcuvisCTE}. Compared with the ACS/WFC, the WFC3/UVIS suffers from more severe CTE loss because the background signal level in WFC3/UVIS images is much lower, making the first few electrons in a pixel particularly susceptible to CTE loss \citep{Anderson2021wfcuvisCTE}. Therefore, using \texttt{\_flc} data corrects for CTE loss and improves the accuracy and reliability of the measurements \citep{Anderson2010EmpiricalCTE}.

In addition, astrometric accuracy is affected by two other effects: image distortion and the breathing effect. The optical system of HST, designed for a sharp PSF, exhibits significant linear and higher-order distortion, causing positional offsets of stars in the images. As a space telescope, the HST optical system is very stable, so the distortion can be computed and corrected \citep{Anderson2022hst1pass}. The HST breathing effect refers to tiny thermal expansions and contractions of the telescope structure due to changes in the thermal environment during its orbit, which cause variations in the focal length and consequently small changes in the PSF. This effect can be corrected by extracting and modelling the PSF from the images \citep{Anderson2022hst1pass}. This study uses the open-source software \texttt{hst1pass} developed by \citet{Anderson2022hst1pass} for source detection, photometry, and astrometry. \texttt{hst1pass} can apply distortion corrections by calling the standard \texttt{STDGDC} geometric distortion correction files, and can also call the standard \texttt{STDPSF} PSF models and perturb them to account for the breathing effect, improving the photometric and astrometric precision.

For each field, our data processing procedure is as follows:
\begin{enumerate}
  \item Run \texttt{hst1pass} on the \texttt{\_flc} files to extract catalogues, performing PSF photometry for unsaturated stars and aperture photometry for saturated stars, obtaining instrumental magnitudes and pixel coordinates. Sources with signal-to-noise ratio (SNR) $>5$ are retained. Instrumental magnitudes are then calibrated to AB magnitudes. Pixel coordinates are mapped to the International Celestial Reference System (ICRS) using the World Coordinate System (WCS) information in the \texttt{\_flc} headers, and astrometric errors are estimated using equation~\eqref{eq:astrometry_error}.
  \item For each epoch, select the F814W \texttt{\_flc} file with the longest exposure as the reference frame, match sources from other F814W files to this reference frame, and reject sources with few matches. For each star, the measured magnitudes and positions from different exposures are weighted averaged by their errors to obtain higher-precision measurements.
  \item If F606W observations exist, match sources to obtain the $m_{\text{F606W}}-m_{\text{F814W}}$ colour.
  \item Match the catalogues from Epoch 1 and Epoch 2. Since there may be systematic astrometric offsets between the two epochs, Legendre polynomials are used to perform the transformation. Tests show that a first-order Legendre polynomial is sufficient to remove background offsets, but in practice a third-order polynomial is used to remove higher-order effects. For the same stars in the two epochs, the coordinates are obtained, and the proper motions and their measurement errors are computed.
  \item Match bright stars in the field with \textit{Gaia} DR3 \citep{GaiaCollaboration2016gaia, GaiaCollaboration2023gaia} catalogue to determine the absolute proper motion correction for each field. The detailed procedure is described in Section~\ref{sec:abs_pm_calibration}.
\end{enumerate}
After the above processing, this study obtains proper motion measurements for about 400,000 stars from the 27 fields.

\subsection{Absolute proper motion calibration}\label{sec:abs_pm_calibration}

For absolute proper motion calibration, this study compares the result of using \textit{Gaia} DR3 with using Optical Gravitational Lensing Experiment \citep[OGLE;][]{Udalski2015OGLEIV} proper motion catalogue from Uranus astrometry system (OGLE/Uranus; Udalski et al., in preparation) as the reference catalogue. Although \textit{Gaia} DR3 and OGLE/Uranus both have proper motion measurements for some relatively faint stars, neither of them can resolve these faint stars well. WFC3/UVIS samples the sky at 0.0395~arcsec pixel$^{-1}$, and its model F814W point spread function (PSF) core has a full width at half maximum (FWHM) of about 0.074~arcsec before pixelation \citep{Marinelli2025WFC3Handbook}. For \textit{Gaia}, the on-board separation limits for equal-brightness pairs are about 0.23~arcsec along scan and 0.70~arcsec across scan \citep{GaiaCollaboration2016gaia,vanLeeuwen2017GaiaPhotometry}, On the other hand, OGLE-IV \citep{Udalski2015OGLEIV} ground-based images have a pixel scale of 0.26 arcsec pixel$^{-1}$ and the OGLE/Uranus subset of these images, about 4000-8000 epochs depending on a field and spanning 15 years, has a seeing-limited stellar PSF FWHM lower than 1.35 arcsec with the median of 1.15 arcsec. Thus, stars separated in the HST images may be blended in the reference catalogues, which causes large differences when comparing the proper motions of faint stars between HST and \textit{Gaia} or OGLE, bringing large uncertainties to the absolute proper motion calibration. This uncertainty becomes apparent as \textit{Gaia} G band reaches 19 or OGLE I band reaches 18. Therefore, a magnitude cut of 19 (18) is applied to both the reference catalogue and the HST catalogue when \textit{Gaia} DR3 (OGLE/Uranus) is used as the reference, and only stars brighter than this cut are used for cross matching and later the absolute proper motion calibration.

We propagate the reference catalogue positions to the HST epoch and cross match the stars. We calculate the dispersion $\sigma_{\Delta\mu}$ of the catalogue-minus-HST differences ($\Delta\mu = \mu_{\text{ref}} - \mu_{\text{HST}}$), and an iterative $4\sigma_{\Delta\mu}$ clipping is then applied to the data to exclude outliers. There are some small spatially varying offsets between the HST proper motions and the reference catalogue, thus a third-order Legendre polynomial is used to calculate the corrections in RA ($\mu_{\alpha}$) and Dec ($\mu_{\delta}$) direction. The resulting corrections are applied to the full HST sample, placing its proper motions on the corresponding reference catalogue system.

Fig.~\ref{fig:pm_gaia_ogle} shows the proper motion comparison in two specific fields between HST and \textit{Gaia} DR3 or OGLE/Uranus after absolute proper motion calibration. The difference in sample sizes between the two fields is partly due to their different spatial areas. In addition, field 18 lies at the edge of one of the chips in the OGLE mosaic camera, reducing the number of available OGLE reference stars in that field by approximately 10--20~per cent. We can see from the standard error of absolute proper motion calibration in Fig.~\ref{fig:pm_gaia_ogle} that OGLE/Uranus results in smaller errors than \textit{Gaia} DR3 in large fields (field 01), but causes much larger errors in small fields (field 18), which cannot be explained solely by the 10--20~per cent reduction in the number of OGLE reference stars. Reducing calibration errors in small fields is particularly important because these fields contain fewer stars and therefore have larger statistical uncertainties. We therefore use \textit{Gaia} DR3 as the reference catalogue for absolute proper motion calibration in this study.

\subsection{Proper motion error analysis}\label{sec:pm_error}

The astrometric error of a star can be estimated from SNR and the FWHM of the PSF \citep{Kozlowski2006kinematics}:
\begin{equation}
  \sigma \simeq \gamma\times\frac{\text{FWHM}}{\text{SNR}},
  \label{eq:astrometry_error}
\end{equation}
where $\gamma=0.6$ if the PSF is perfectly Gaussian. Tests in this study show that equation~\eqref{eq:astrometry_error} describes the distribution of astrometric errors well.

The pixel scale of the WFC3/UVIS camera is about 40~mas pixel$^{-1}$, and the FWHM of the PSF in the F814W band is about 1.8 pixels. Therefore, according to equation~\eqref{eq:astrometry_error}, a star with SNR = 10 in both epochs would have an astrometric error of about 4 mas per epoch; considering a 10-year baseline, this yields a proper motion precision of $\sim 0.6$~mas yr$^{-1}$. For stars with higher SNR, the precision can be further improved, reaching errors $< 0.1$~mas yr$^{-1}$. Typical bulge stars have velocities of about 100~km s$^{-1}$ at a distance of about 8 kpc, corresponding to proper motions of about 2.6~mas yr$^{-1}$. Therefore, the proper motion precision achieved in this study is well suited for the subsequent analysis.

The absolute proper motion calibration using \textit{Gaia} data also introduces errors. After matching with the \textit{Gaia} catalogue, the difference between the \textit{Gaia} proper motions and the HST proper motions $\Delta\mu = \mu_{\text{ref}} - \mu_{\text{HST}}$ is calculated. We then use
\begin{equation}
  \sigma_{\text{abs}} = \frac{\sigma_{\Delta\mu}}{\sqrt N}
  \label{eq:abs_calib_error}
\end{equation}
to calculate the absolute calibration error, where $\sigma_{\Delta\mu}$ is the standard deviation of $\Delta\mu$, and $N$ is the number of calibration stars in the field. In this study, the absolute calibration error is controlled to be $\lesssim0.1$~mas yr$^{-1}$ for each field.

% !TEX root = ../mnras_template.tex

\section{Statistical Analyses}\label{chap:03statistic_method}

\begin{figure}
  \centering
  \includegraphics[width=0.8\linewidth]{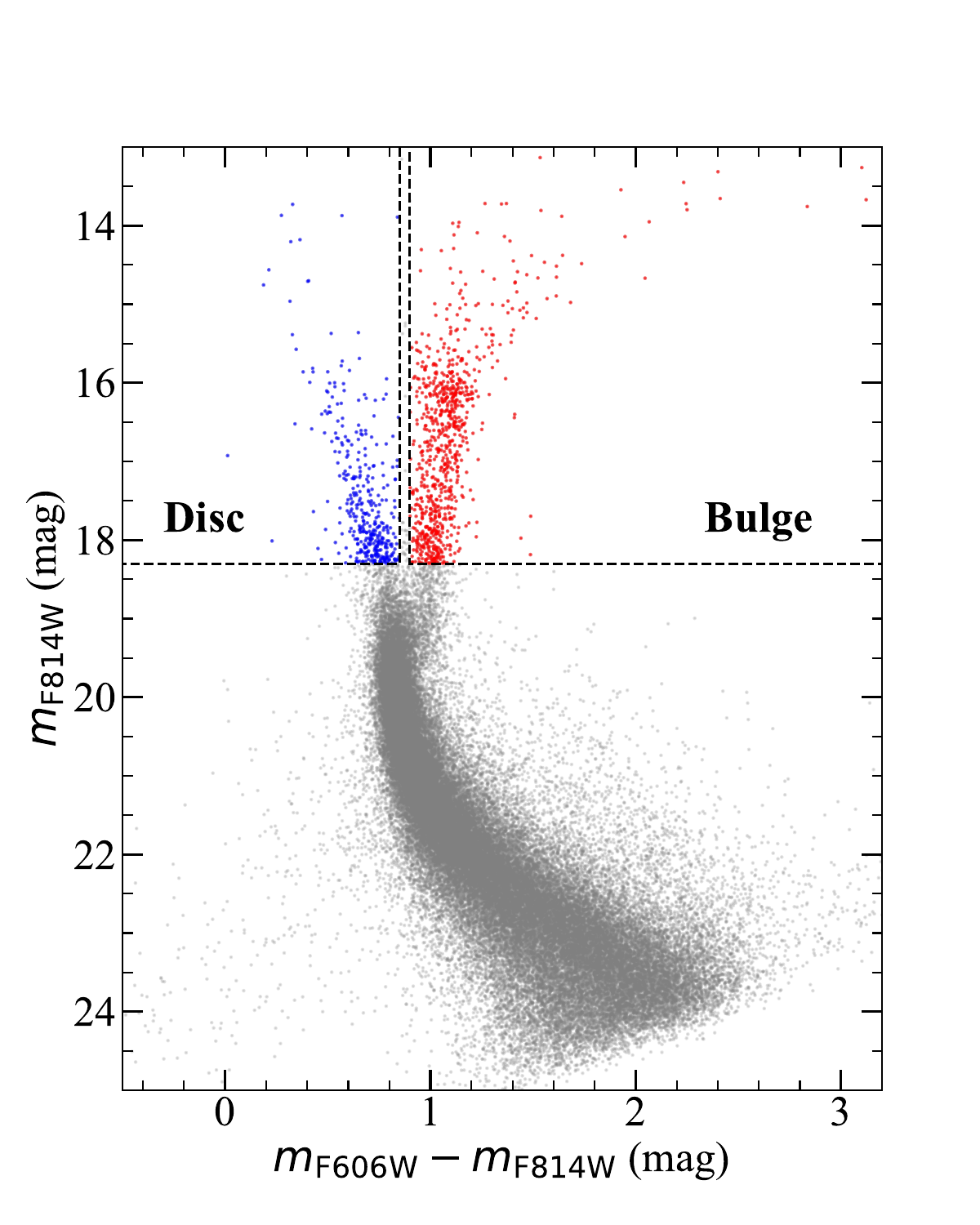}
  \caption{CMD for Field 15 ($l = 1.16^{\circ}$, $b = -2.73^{\circ}$). This CMD is constructed from the Epoch 1 data of this field and contains 100,190 stars. The red data points in the upper right are mainly bulge RC/RGB stars, while the blue data points in the upper left are mainly foreground disc MS stars. The black dashed lines indicate the colour and magnitude ranges of RC/RGB stars and foreground MS stars, respectively.}
  \label{fig:CMD}
\end{figure}

\begin{figure}
  \centering
  \includegraphics[width=1\linewidth]{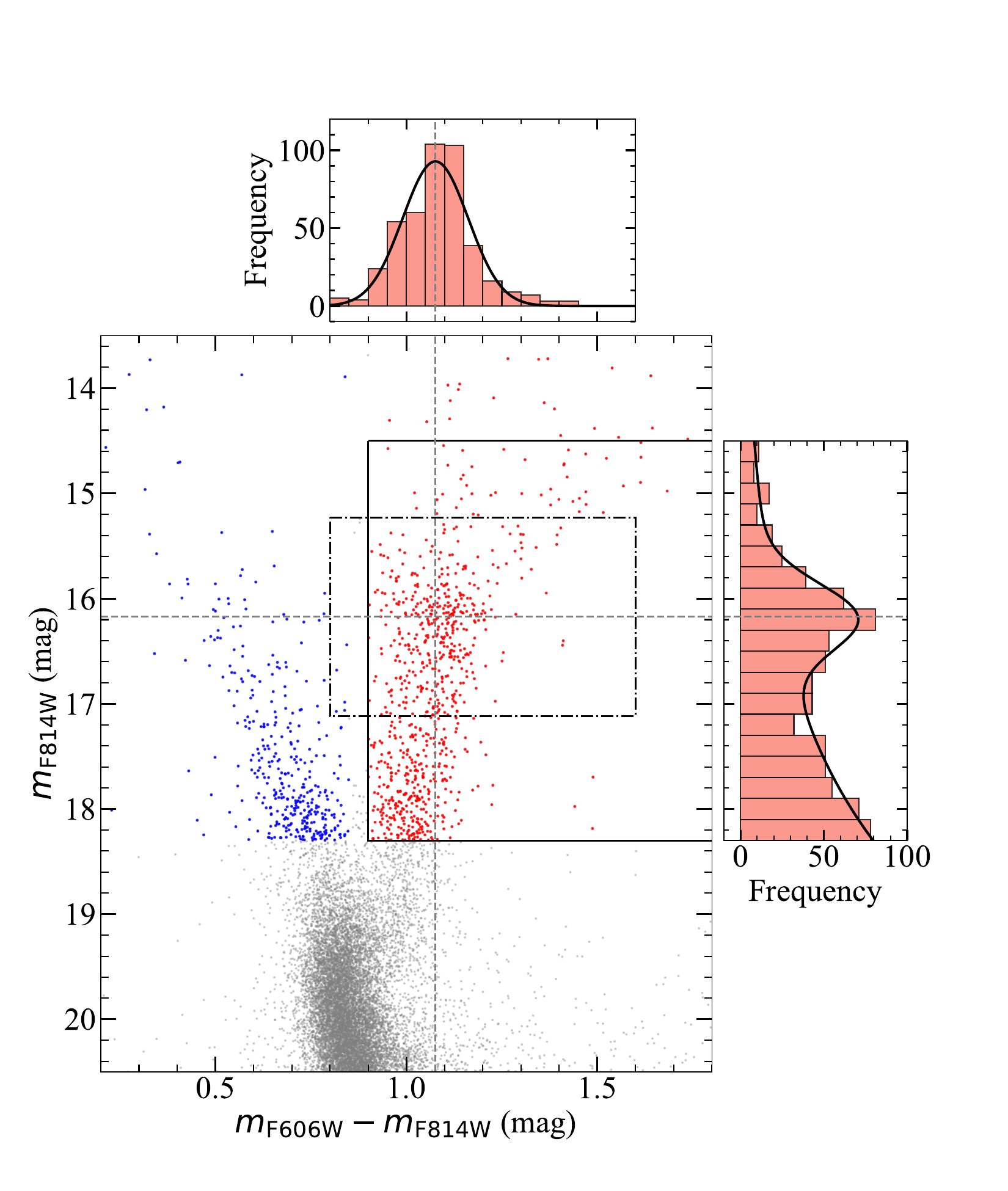}
  \caption{The selected regions in the CMD for the fits of RC star magnitudes (black solid line) and colours (black dash-dotted line). The two panels show the distributions and fitted curves for colour (top) and magnitude (right) within the selected regions. The grey dashed lines indicate the fitted magnitude and colour of the RC stars.}
  \label{fig:Extinction_Color_Fit}
\end{figure}

\subsection{Mean proper motion, dispersion, anisotropy, and covariance}

After obtaining the absolute proper motions of stars in each field, we transform the proper motions from the equatorial coordinate system to the Galactic coordinate system, obtaining the position of each field in $(l,b)$ coordinates, and the proper motions $\mu_l$ and $\mu_b$ for each star. Because of differences in extinction, exposure times, and other factors across fields, we need to apply a magnitude cut to the stellar sample to reduce systematic errors when performing statistical analysis. For each field, we select stars with magnitudes $18.5 \le m_{\text{F814W}} \le 23$, and compute the mean absolute proper motions $\langle\mu_l\rangle$ and $\langle\mu_b\rangle$, as well as the proper motion dispersions $\sigma_{l}$ and $\sigma_{b}$. The proper motion dispersions are computed following \citet{MotionsinPleiades1970}:
\begin{equation}
  \sigma_{\mu} = \sqrt{\frac{1}{N-1}\sum_{i=1}^N (\mu_i - \langle\mu\rangle)^2 - \frac{1}{N}\sum_{i=1}^N \delta_{\mu_i}^2},
\end{equation}
where $\mu_i$ is the proper motion of the $i$-th star, $\langle\mu\rangle$ is the mean proper motion, $\delta_{\mu_i}$ is the proper motion measurement error of the $i$-th star, and $N$ is the number of selected stars. This formula assumes that the observed dispersion is the quadrature sum of the intrinsic dispersion and the measurement errors, so subtracting the contribution of measurement errors yields a dispersion that better reflects the true velocity distribution of the stars. After obtaining $\sigma_{l}$ and $\sigma_{b}$, following \citet{Kozlowski2006kinematics}, we compute the ratio $\sigma_{l}/\sigma_{b}$ to characterise the degree of velocity anisotropy of bulge stars, and the normalised covariance $C_{lb} := \sigma_{lb}/(\sigma_{l}\sigma_{b})$ to characterise the degree of asymmetry in the velocity distribution. To reduce the impact of outliers, we remove stars with proper motions deviating from the mean by more than $3\sigma$ before computing the above statistics.

\subsection{Error estimation}

To estimate the errors of the above statistics, we adopt the bootstrap method. The most critical parameter in the bootstrap method is the number of resampling iterations (steps). If too few steps are chosen, the error estimates may not converge; if too many, the computational cost increases significantly. To determine the optimal number of bootstrap steps, we select four fields with different numbers of stars and analyse the variations of the statistical results and their standard errors with different numbers of bootstrap steps. The four fields have 1,184, 2,892, 10,108, and 43,924 stars, respectively. We find that the standard errors converge more slowly than the statistics themselves, and fields with more stars converge faster. When the number of bootstrap steps reaches $\sim 500$, the statistical results have essentially converged, and the standard errors also become stable. To ensure the reliability of the statistical results, we adopt 3,000 steps as the number of bootstrap resampling iterations in this study.

\subsection{Colour–magnitude diagrams}\label{sec:CMD}

Among the 27 fields, 16 have photometric data in both the F606W and F814W bands, allowing us to construct colour–magnitude diagrams (CMDs) for each field using the colour $m_{\text{F606W}} - m_{\text{F814W}}$. Fig.~\ref{fig:CMD} shows the CMD for Field 15.

Along the line of sight towards the bulge, we observe not only bulge stars but also a large number of foreground disc stars, which contaminate the kinematic analysis of bulge stars. To isolate the kinematic properties of bulge stars more cleanly, we need to distinguish bulge stars from disc stars among the bright stars. In Fig.~\ref{fig:CMD}, the main-sequence turn-off of bulge stars can be seen at $m_{\text{F814W}}\sim 19$, and above the turn-off two branches can be distinguished. The red data points in the upper right of the CMD are mainly bulge red clump (RC) and red giant branch (RGB) stars, while the bluer branch in the upper left consists of a large number of foreground disc main-sequence (MS) stars. Therefore, by analysing the CMD morphology and applying colour cuts, we can separate bulge RC/RGB stars from foreground disc stars.

\subsection{Extinction and reddening of bulge stars}\label{sec:extinction_reddening}

Furthermore, due to the severe interstellar extinction towards the bulge, interstellar dust not only dims the stars but also reddens their colours. RC stars are particularly prominent in this direction and have similar magnitudes and colours, making them concentrated in the CMD and often used for distance estimation and extinction correction. Following the method of \citet{Nataf2013ReddeningExtinction}, we fit the magnitude distribution of RC/RGB stars in the CMD using the function
\begin{equation}
  \begin{split}
    N(I) =& A \exp[B(I - I_{\text{RC}})] + \frac{N_{\text{RC}}}{\sqrt{2\pi}\sigma_{\text{RC}}} \exp\left[-\frac{(I - I_{\text{RC}})^2}{2\sigma_{\text{RC}}^2}\right] \\
    &+ \frac{N_{\text{RGBB}}}{\sqrt{2\pi}\sigma_{\text{RGBB}}} \exp\left[-\frac{(I - I_{\text{RGBB}})^2}{2\sigma_{\text{RGBB}}^2}\right] \\
    &+ \frac{N_{\text{AGBB}}}{\sqrt{2\pi}\sigma_{\text{AGBB}}} \exp\left[-\frac{(I - I_{\text{AGBB}})^2}{2\sigma_{\text{AGBB}}^2}\right],
  \end{split}
  \label{eq:extinction_fit_Nataf2013}
\end{equation}
where $I$ is the I-band magnitude, $N$ is the number of stars, and $\sigma$ is the standard deviation of the magnitude distribution. This function assumes that the background star counts follow an exponential distribution with magnitude, and in addition to RC stars, it also accounts for two other minor but similarly concentrated populations, namely the red giant branch bump (RGBB) and the asymptotic giant branch bump (AGBB). However, the fields in this study are not large enough to have enough bright stars to support the triple-Gaussian fit of equation~\eqref{eq:extinction_fit_Nataf2013}, so we adopt a simpler fitting function that includes only the exponential background term and the Gaussian term for RC stars:
\begin{equation}
  \begin{split}
    N(m) =& A \exp[B(m - m_{\text{RC}})] \\
    &+ \frac{N_{\text{RC}}}{\sqrt{2\pi}\sigma_{m_\text{RC}}} \exp\left[-\frac{(m - m_{\text{RC}})^2}{2\sigma_{m_\text{RC}}^2}\right],
  \end{split}
  \label{eq:extinction_fit_RC}
\end{equation}
where $m$ is the F814W magnitude. For the colour distribution, we use a single Gaussian without a background component:
\begin{equation}
  N(C) = \frac{N_{\text{RC}}}{\sqrt{2\pi}\sigma_{C_{\text{RC}}}} \exp\left[-\frac{(C - C_{\text{RC}})^2}{2\sigma_{C_{\text{RC}}}^2}\right],
  \label{eq:reddening_fit_RC}
\end{equation}
where $C = m_{\text{F606W}} - m_{\text{F814W}}$ is the colour. From the fits of equations~\eqref{eq:extinction_fit_RC} and \eqref{eq:reddening_fit_RC}, we obtain the mean magnitude $m_{\text{RC}}$ and mean colour $C_{\text{RC}}$ of the RC stars. Fig.~\ref{fig:Extinction_Color_Fit} shows the fitting results for Field 15, for which $m_{\text{RC}} = 16.173 \pm 0.036$ and $C_{\text{RC}} = 1.076 \pm 0.006$. By performing the above fits for all 16 fields, we obtain the distributions of $m_{\text{RC}}$ and $C_{\text{RC}}$ for each field. Since the absolute magnitude and colour of RC stars are essentially the same across different fields, we can calibrate the measured $m_{\text{RC}}$ and $C_{\text{RC}}$ to a reference value, thereby deriving the reddening and extinction along the line of sight for each field. This provides a calibration reference for the kinematic analysis of stars in different magnitude and colour ranges. In this work, we choose the reference values to be $m_{\text{RC}} = 16$ and $C_{\text{RC}} = 1$, and we denote the extinction-corrected magnitude as $m'_{\text{F814W}}$ and reddening-corrected colour as $C'_{\text{F606W}-\text{F814W}}$.

% !TEX root = ../mnras_template.tex

\begin{figure*}
  \centering
  \includegraphics[width=0.8\linewidth]{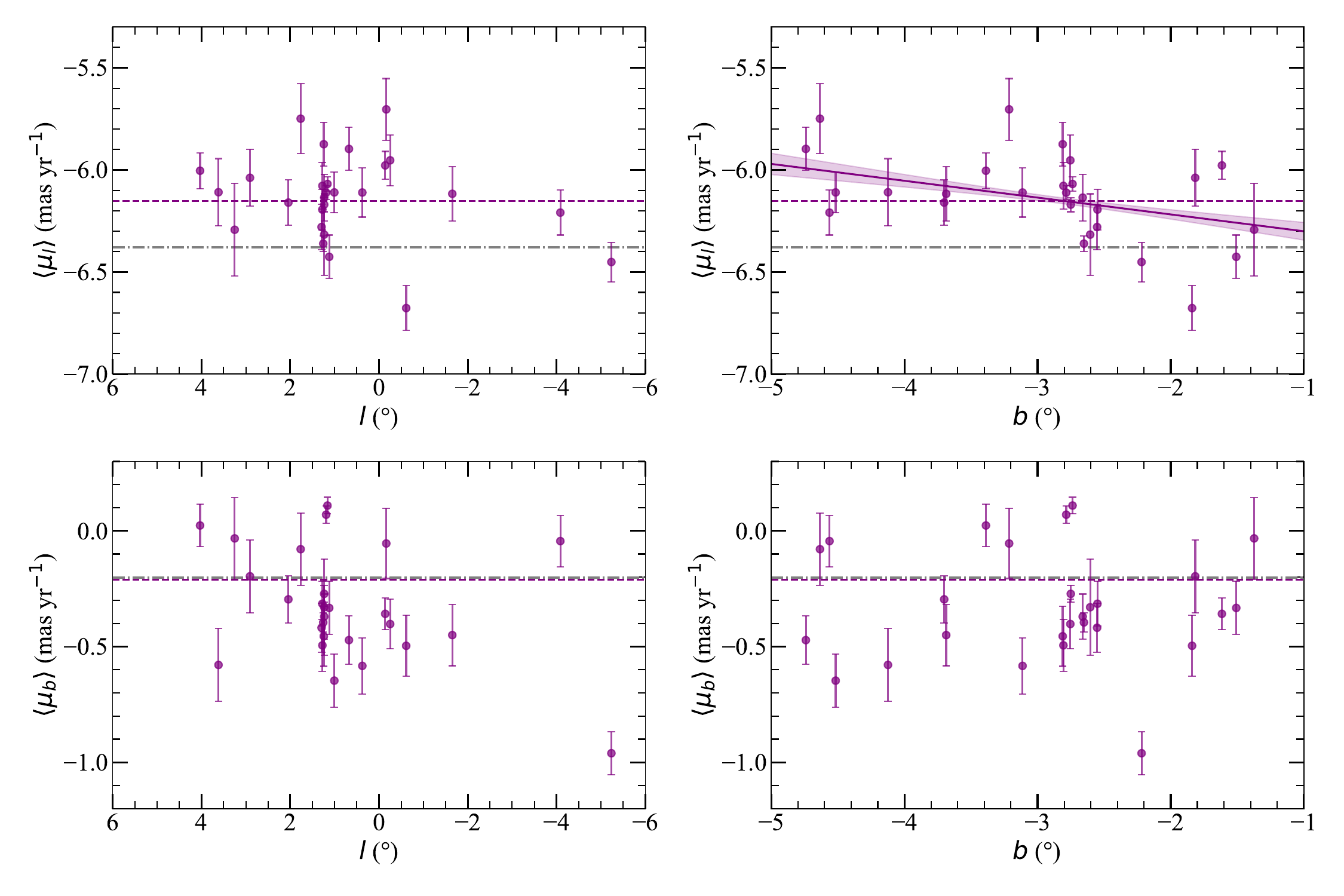}
  \caption{Distributions of the mean proper motions $\langle\mu_{l}\rangle$ and $\langle\mu_{b}\rangle$ as functions of $l$ and $b$. The purple dashed lines show the average values of the mean proper motions, and the grey dash-dotted lines show the proper motion of Sgr A* measured by \citet{Reid2004SgrAProperMotion}. In the upper-right panels, the purple solid line and shaded region show the linear fit and $1\sigma$ error band.}
  \label{fig:mean_pm}
\end{figure*}

\begin{figure*}
  \centering
  \includegraphics[width=0.8\linewidth]{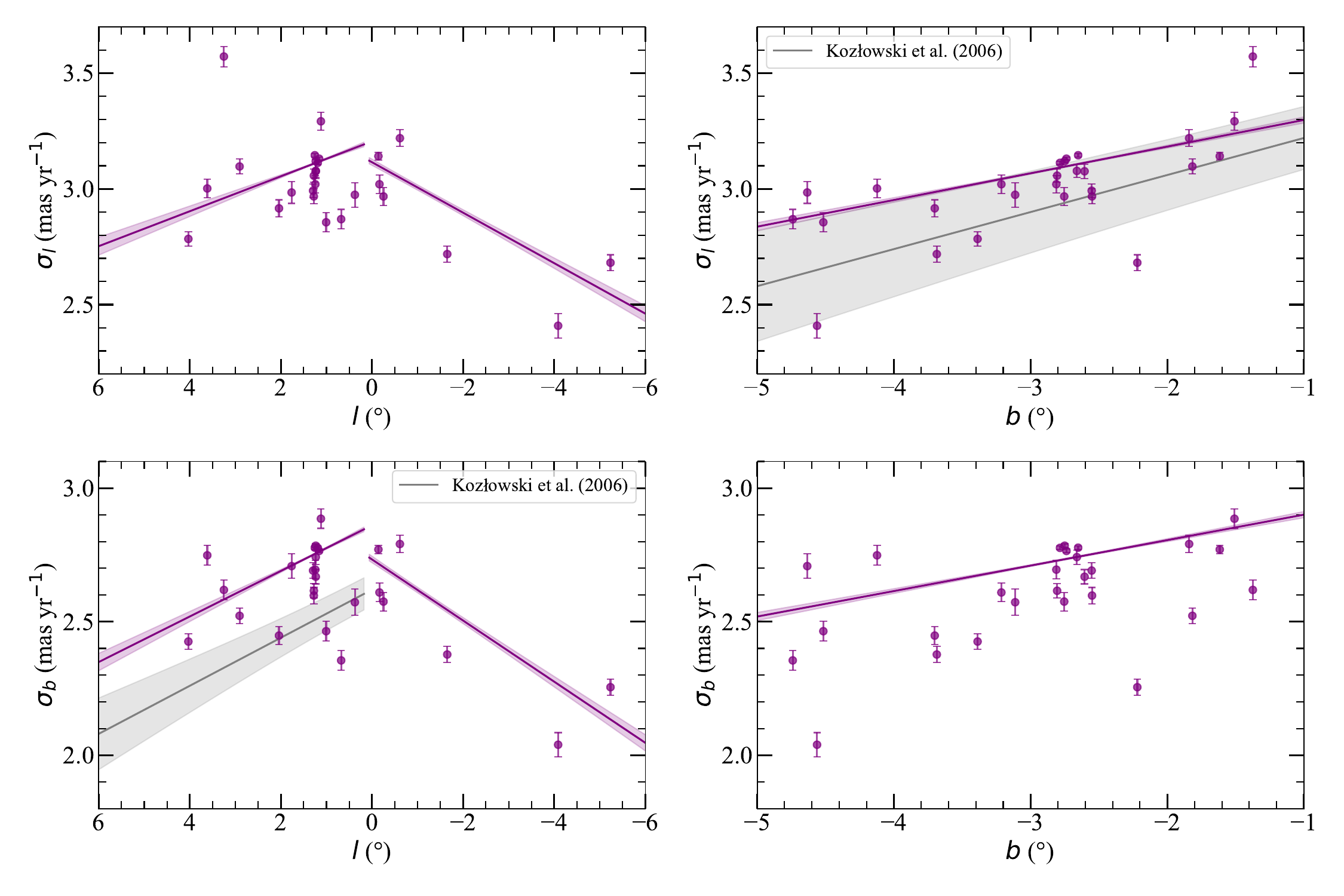}
  \caption{Distributions of the proper motion dispersions $\sigma_{l}$ and $\sigma_{b}$ as functions of $l$ and $b$, along with linear fits. The purple solid lines and shaded regions indicate the linear fits and their $1\sigma$ error bands. In the left panels, the fits for $l > 0^{\circ}$ and $l < 0^{\circ}$ are performed separately. The outlying point with the largest $\sigma_l$ is excluded from the $\sigma_l$ fits. The grey solid lines and shaded regions show the trends of proper motion dispersions obtained by \citet{Kozlowski2006kinematics} near Baade's Window.}
  \label{fig:pm_dispersion}
\end{figure*}

\begin{figure*}
  \centering
  \includegraphics[width=0.8\linewidth]{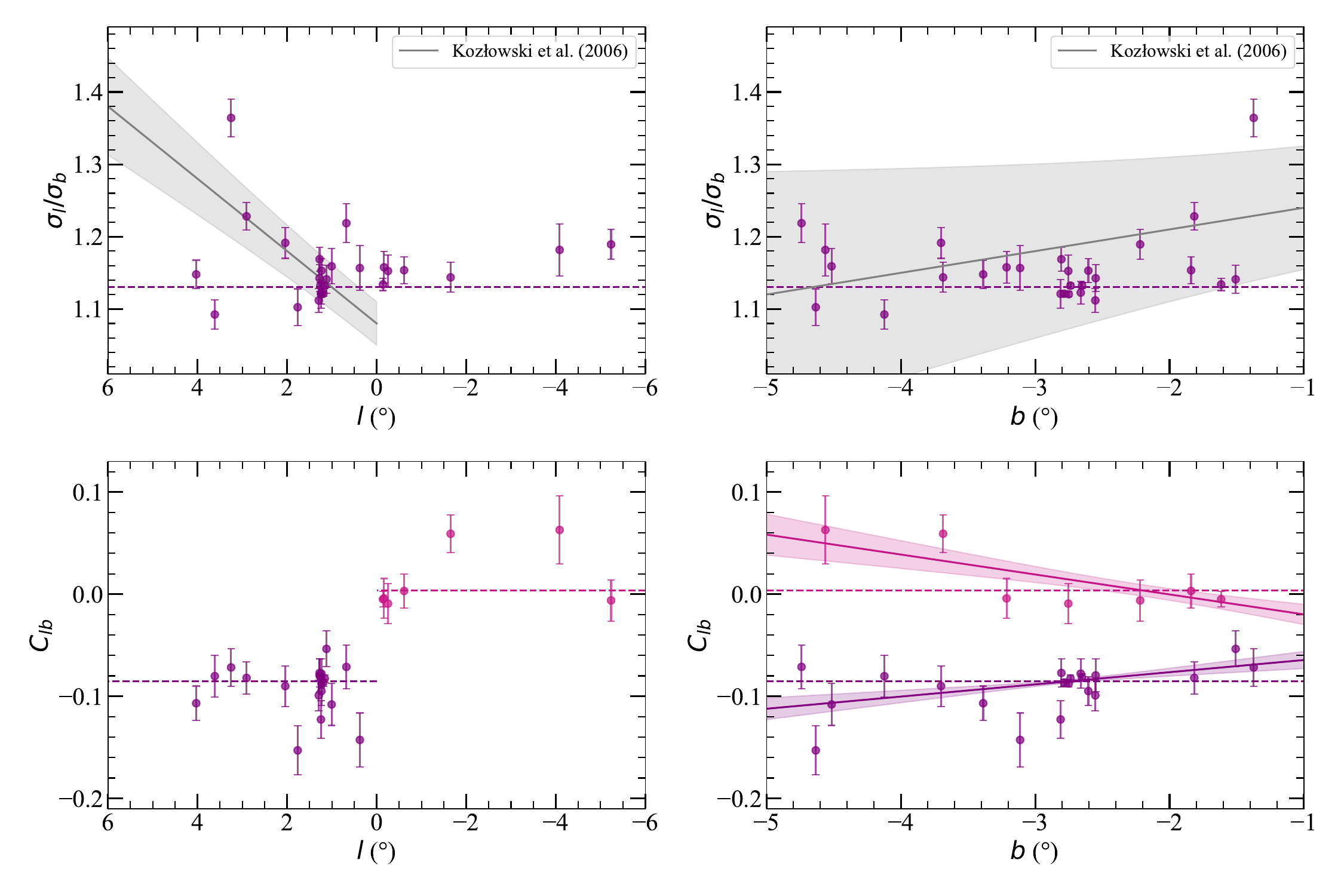}
  \caption{Distributions of $\sigma_l / \sigma_b$ (top panels) and the normalised covariance $C_{lb}$ (bottom panels) of bulge stellar proper motion dispersions as functions of Galactic coordinates $l$ and $b$. In the top panels, the purple dashed lines show the mean value, with the outlying largest $\sigma_l/\sigma_b$ point excluded. The grey solid lines and shaded regions show the trends from \citet{Kozlowski2006kinematics}. In the bottom panels, the data are separated into $l > 0^{\circ}$ (purple) and $l < 0^{\circ}$ (magenta), with their respective mean values (dashed lines); the lower-right panel shows linear fits to the two groups (solid lines and shaded regions).}
  \label{fig:pm_rsig_cov}
\end{figure*}

\section{Results}\label{chap:04results}

\subsection{Mean proper motions}

Fig.~\ref{fig:mean_pm} shows the spatial distributions of the mean proper motions $\langle\mu_l\rangle$ and $\langle\mu_b\rangle$ for the 27 fields. Averaging over all fields gives $\langle\mu_l\rangle = -6.152 \pm 0.036$~mas yr$^{-1}$ and $\langle\mu_b\rangle = -0.212 \pm 0.054$~mas yr$^{-1}$. \citet{Reid2004SgrAProperMotion} measured the proper motion of the supermassive black hole Sgr A* at the Galactic centre using the Very Long Baseline Array (VLBA), obtaining $\mu_l = -6.379 \pm 0.026$~mas yr$^{-1}$ and $\mu_b = -0.202 \pm 0.019$~mas yr$^{-1}$. The proper motion of Sgr A* can be taken as representative of the bulge centre, mainly reflecting the motion of the Solar System relative to the bulge centre. Our measured $\langle\mu_b\rangle$ is consistent with that of \citet{Reid2004SgrAProperMotion}, while there is a significant difference in $\langle\mu_l\rangle$. This discrepancy is likely because our mean proper motions are averaged over all stars with $18.5 < m_{\text{F814W}} < 23$ in the 27 fields, so spatial offsets and contamination by disc stars can affect our results. As discussed in Section~\ref{sec:NonGaussianity}, disc stars move mainly along the Galactic plane and more positively in $l$ direction, making $\langle\mu_l\rangle$ more susceptible to disc contamination, while $\langle\mu_b\rangle$ is less affected, which explains why the mean proper motion in $b$ agrees better with that of Sgr A*. \citet{Sanders2019VVVGaiaKinematics} and \citet{Steven2022BulgePMRotationCurves} found that even inside the bulge, $\mu_l$ also shows a more positive value at closer distance, while $\mu_b$ is nearly constant. This strong distance dependence shows that the line-of-sight selection can also contribute to the offset of our mean $\langle\mu_l\rangle$ from Sgr A*.

From Fig.~\ref{fig:mean_pm}, we also see a trend of $\langle\mu_l\rangle$ with $b$: fields closer to the Galactic plane ($b=0^{\circ}$) tend to have more negative mean proper motions. A linear fit gives
\begin{equation}
  \langle\mu_l\rangle ~(b) = (-0.082 \pm 0.023)\times b + (-6.383 \pm 0.065),
  \label{eq:mean_pm_fit_mul_b}
\end{equation}
where $b$ is in degrees and $\langle\mu_l\rangle$ in mas yr$^{-1}$. According to equation~\eqref{eq:mean_pm_fit_mul_b}, extrapolating the fit to the Galactic plane ($b=0^{\circ}$) gives $\langle\mu_l\rangle = -6.383 \pm 0.065$~mas yr$^{-1}$, which is consistent with $\mu_l = -6.379 \pm 0.026$~mas yr$^{-1}$ from \citet{Reid2004SgrAProperMotion}. This indicates that the trend we observe can, to some extent, reflect the true spatial variation of the proper motions.

\subsection{Proper motion dispersions}

Fig.~\ref{fig:pm_dispersion} shows the spatial distributions of the proper motion dispersions $\sigma_{l}$ and $\sigma_{b}$. Compared with the mean proper motions, the dispersions show clearer gradients: fields closer to the Galactic centre ($l=0^{\circ}, b=0^{\circ}$) have larger dispersions. Since our fields cover both $l>0^{\circ}$ and $l<0^{\circ}$, we perform separate linear fits for the two sides. Furthermore, as seen in the upper panels of Fig.~\ref{fig:pm_dispersion}, there is an outlier with an unusually large $\sigma_l$, which we exclude from the fit. The resulting fits are:
\begin{subnumcases}{\sigma_{l}~(l)=}
  (-0.076 \pm 0.008)\times l + (3.206 \pm 0.011), & $l>0$, \label{eq:pm_dispersion_fit_mul_l_1} \\
  (0.109 \pm 0.007)\times l + (3.115 \pm 0.013), & $l<0$, \label{eq:pm_dispersion_fit_mul_l_2}
\end{subnumcases}
\begin{subnumcases}{\sigma_{b}~(l)=}
  (-0.085 \pm 0.007)\times l + (2.861 \pm 0.010), & $l>0$, \label{eq:pm_dispersion_fit_mub_l_1} \\
  (0.114 \pm 0.006)\times l + (2.733 \pm 0.012), & $l<0$, \label{eq:pm_dispersion_fit_mub_l_2}
\end{subnumcases}
\begin{equation}
  \sigma_{l}~(b) = (0.115 \pm 0.007)\times b + (3.413 \pm 0.021),
  \label{eq:pm_dispersion_fit_mul_b}
\end{equation}
\begin{equation}
  \sigma_{b}~(b) = (0.095 \pm 0.007)\times b + (2.996 \pm 0.018),
  \label{eq:pm_dispersion_fit_mub_b}
\end{equation}
where $l$ and $b$ are in degrees, and $\sigma_{l}$ and $\sigma_{b}$ are in mas yr$^{-1}$.

\citet{Kozlowski2006kinematics} analysed the proper motion dispersions of bulge stars in 37 fields near Baade's Window and obtained linear trends over the ranges $0.5^{\circ} < l < 5.5^{\circ}$ and $-4.5^{\circ} < b < -2^{\circ}$: $\sigma_{l}~(b) = (0.16 \pm 0.04)\times b + (3.38 \pm 0.13)$~mas yr$^{-1}$ and $\sigma_{b}~(l) = (-0.09 \pm 0.02)\times l + (2.62 \pm 0.06)$~mas yr$^{-1}$ (see Fig.~\ref{fig:pm_dispersion}). Within the overlapping range, our slopes from equations~\eqref{eq:pm_dispersion_fit_mul_b} and \eqref{eq:pm_dispersion_fit_mub_l_1} are broadly consistent with theirs, but the intercepts differ somewhat. Meanwhile, equations~\eqref{eq:pm_dispersion_fit_mul_l_1} and \eqref{eq:pm_dispersion_fit_mub_b} show clearer trends that were not reported by \citet{Kozlowski2006kinematics}. These differences may arise from different magnitude ranges used: our study uses the more sensitive WFC3/UVIS camera on HST, which reaches fainter stars, with $18.5 < m_{F814W} < 23$, while \citet{Kozlowski2006kinematics} used the Planetary Camera (PC) of WFPC2 and the High Resolution Channel (HRC) of ACS and selected $18.0 < m_{F814W} < 21.5$. Moreover, our larger sample size yields smaller errors.

Our data cover both sides of the bulge, allowing us to investigate asymmetries in the dispersions. While both sides show increasing dispersions towards the centre, the slopes differ. From equations~\eqref{eq:pm_dispersion_fit_mul_l_1} and \eqref{eq:pm_dispersion_fit_mul_l_2}, the ratio of slopes is $r_{\sigma_l} = \left|k_{\sigma_l|l>0}~/~k_{\sigma_l|l<0}\right| = 0.70 \pm 0.09$; from equations~\eqref{eq:pm_dispersion_fit_mub_l_1} and \eqref{eq:pm_dispersion_fit_mub_l_2}, $r_{\sigma_b} = \left|k_{\sigma_b|l>0}~/~k_{\sigma_b|l<0}\right| = 0.75 \pm 0.07$. Both ratios are significantly less than 1, indicating that the kinematic state is asymmetric between the two sides, with a steeper gradient on the $l<0$ side. This asymmetry may be caused by non-axisymmetric structures such as the bar.

\citet{Wang2013SchwarzschildModelBar} constructed a dynamical model of the Galactic bar using the potential from $N$-body simulations and Schwarzschild method, fitting to BRAVA survey data, and obtained a bar pattern speed $\Omega_{\rm p}=40\ \mathrm{km\ s^{-1}\ kpc^{-1}}$ and viewing angle $\theta_{\rm bar}=45^\circ$. Compared with our measured dispersions, the model of \citet{Wang2013SchwarzschildModelBar} predicts slightly lower $\sigma_b$ but significantly higher $\sigma_l$ than our results.

\subsection{Anisotropy and covariance}

Fig.~\ref{fig:pm_rsig_cov} shows the spatial distributions of the anisotropy $\sigma_l / \sigma_b$ and the normalised covariance $C_{lb}$ of the bulge stellar proper motion dispersions. The mean $\sigma_l / \sigma_b$ is $1.131 \pm 0.005$, with all data points $>1.0$. This is in general agreement with \citet{Kozlowski2006kinematics}, who gave a mean $\sigma_l / \sigma_b$ between $1.1 \sim 1.2$, indicating that bulge stars are significantly anisotropic, with larger dispersion in the $l$ direction. \citet{Kozlowski2006kinematics} found trends of $\sigma_l/\sigma_b$ with $l$ and $b$ (see Fig.~\ref{fig:pm_rsig_cov}), but with large uncertainties, while we do not find such trends in our data. For the normalised covariance $C_{lb}$, the values on the $l>0$ and $l<0$ sides show a clear difference, so we treat them separately. The mean covariance for $l>0$ is $-0.085 \pm 0.003$, also consistent with \citet{Kozlowski2006kinematics}, who gave $C_{lb}\sim -0.1$, indicating a significant negative correlation between the $l$ and $b$ velocity components on this side. For $l<0$, the mean covariance is $0.004 \pm 0.011$, showing a weaker correlation and even positive correlations in some fields. The absolute difference between the two mean values further highlights the kinematic asymmetry of the bulge. The integrated VIRAC maps of \citet{Clarke2019VIRAC3D} show $\sigma_l/\sigma_b$ slightly above 1.1 along the minor axis and a quadrupolar proper-motion correlation of order 0.1 that is stronger at positive $l$, which is also interpreted as a consequence of the viewing angle of the bar. These patterns agree with our mean anisotropy and the stronger covariance on the $l>0$ side.

We find that $C_{lb}$ shows a trend with $b$, and the trend is opposite for $l>0$ and $l<0$. Separate linear fits give:
\begin{subnumcases}{C_{lb}=}
  (0.012 \pm 0.005)\times b - (0.053 \pm 0.013), & $l>0$, \label{eq:cov_fit_Clb_b_1} \\
  (-0.020 \pm 0.007)\times b - (0.039 \pm 0.016), & $l<0$, \label{eq:cov_fit_C_lb_b_2}
\end{subnumcases}
where $b$ is in degrees. According to this trend, the asymmetry in $C_{lb}$ between the two sides increases with distance from the Galactic plane ($b=0^{\circ}$). However, given the limited data on the $l<0$ side, the robustness of this trend requires further verification.

\begin{figure}
  \centering
  \includegraphics[width=1\linewidth]{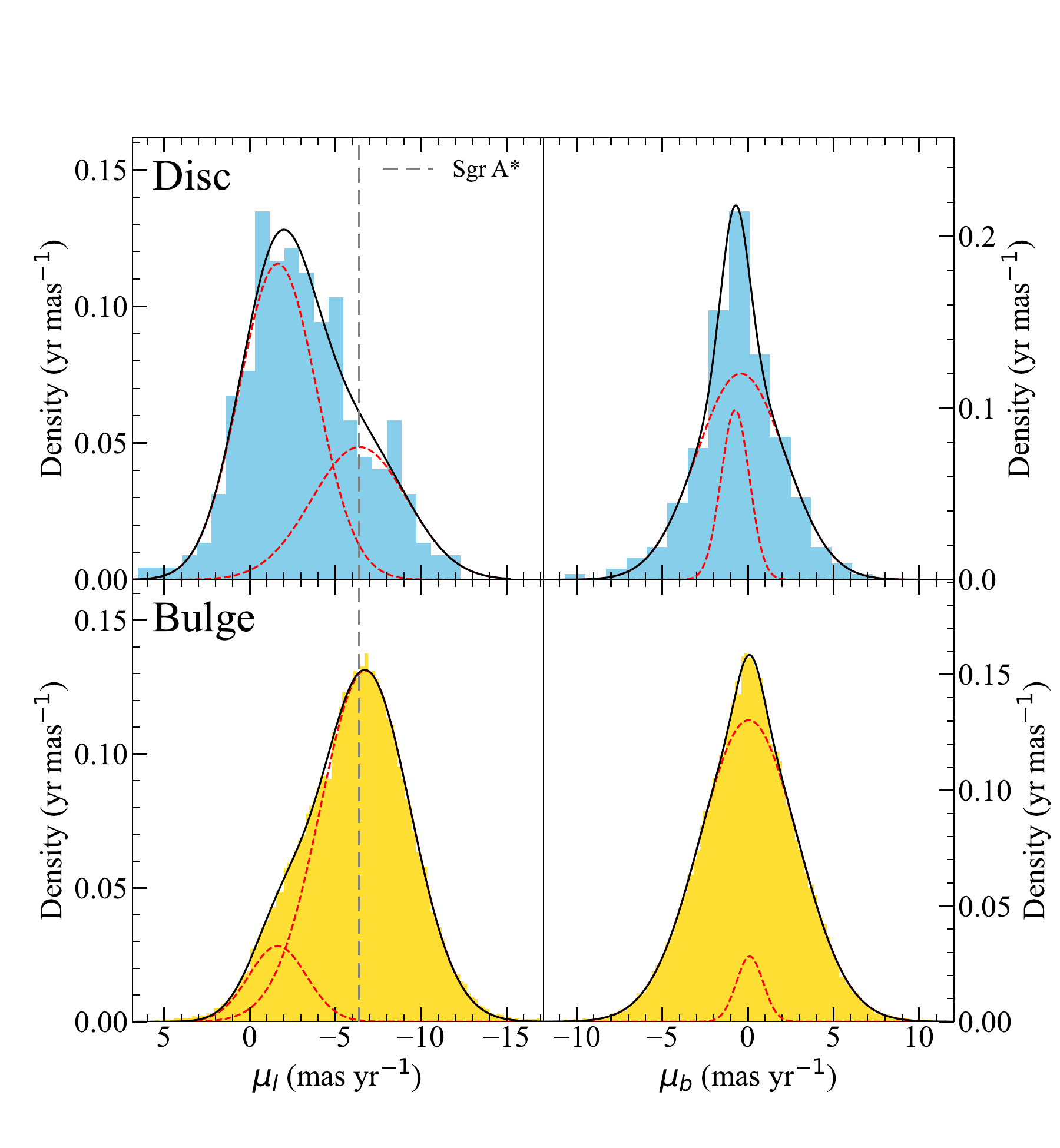}
  \caption{Absolute proper motion distributions of disc stars (upper panels) and bulge stars (lower panels) in Field 14 ($l = 1.19^{\circ}$, $b = -2.78^{\circ}$). The disc stars are the blue bright stars in the upper-left branch of the CMD (260 in total). The remaining stars are classified as bulge stars (61,205 in total). Each distribution is fitted with a double Gaussian. The two red dashed lines indicate the two individual Gaussians, and the black solid line shows the combined fit. The $\mu_l$ of Sgr A* \citep{Reid2004SgrAProperMotion} is marked with a vertical grey dashed line.}
  \label{fig:pm_distribution_27}
\end{figure}

\begin{figure*}
  \centering
  \includegraphics[width=0.8\linewidth]{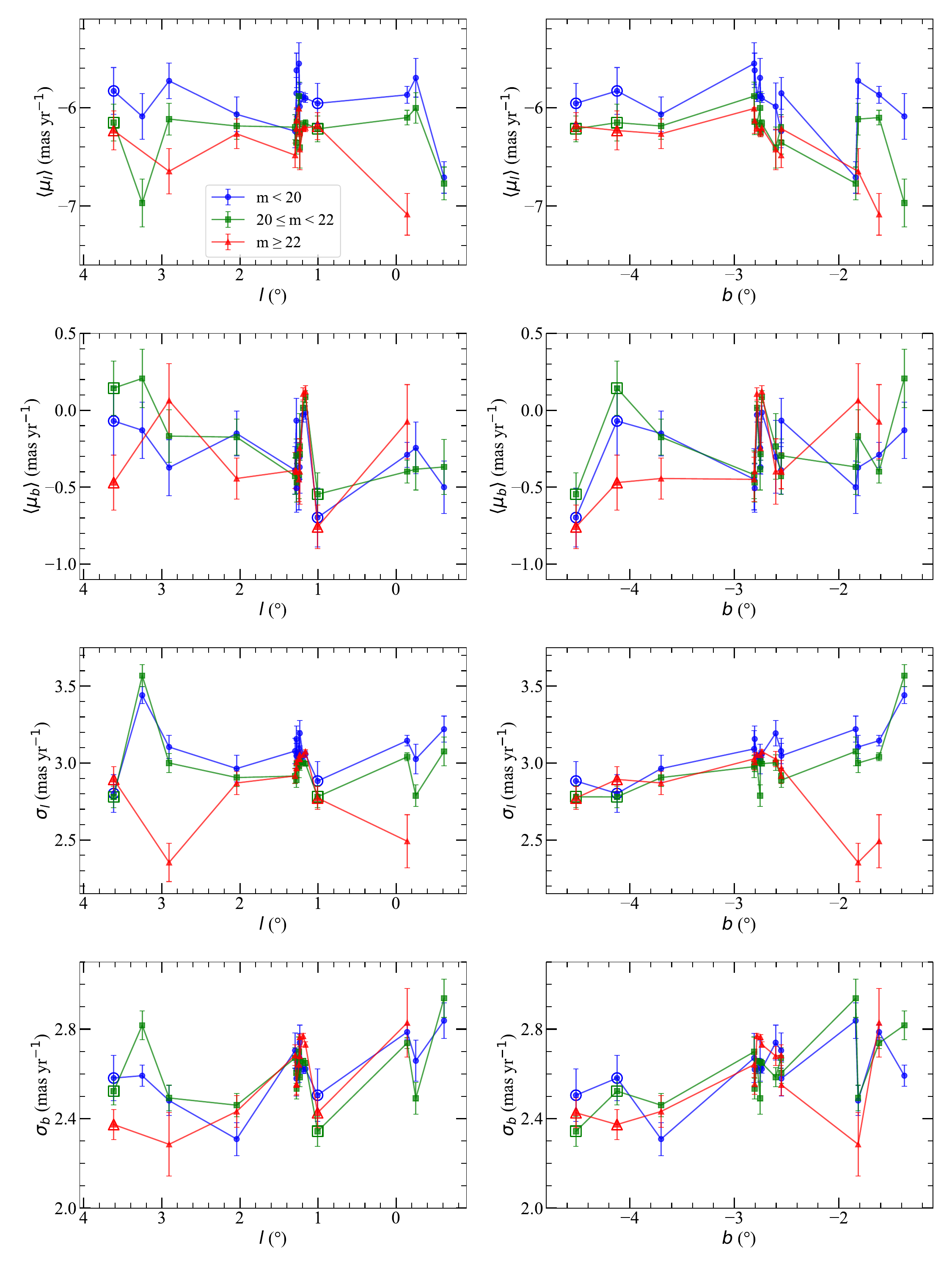}
  \caption{Extinction-corrected magnitude $m'_{\text{F814W}}$ (defined at the end of Section~\ref{sec:extinction_reddening}) ranges: $m'_{\text{F814W}} < 20$ (blue circles), $20 \le m'_{\text{F814W}} < 22$ (green squares), $m'_{\text{F814W}} \ge 22$ (red triangles). Fields 26 and 27 (marked with borders) are not corrected for extinction due to the small number of bright stars. Data points with fewer than 100 stars are not included.}
  \label{fig:pm_mag}
\end{figure*}

\begin{figure*}
  \centering
  \includegraphics[width=0.8\linewidth]{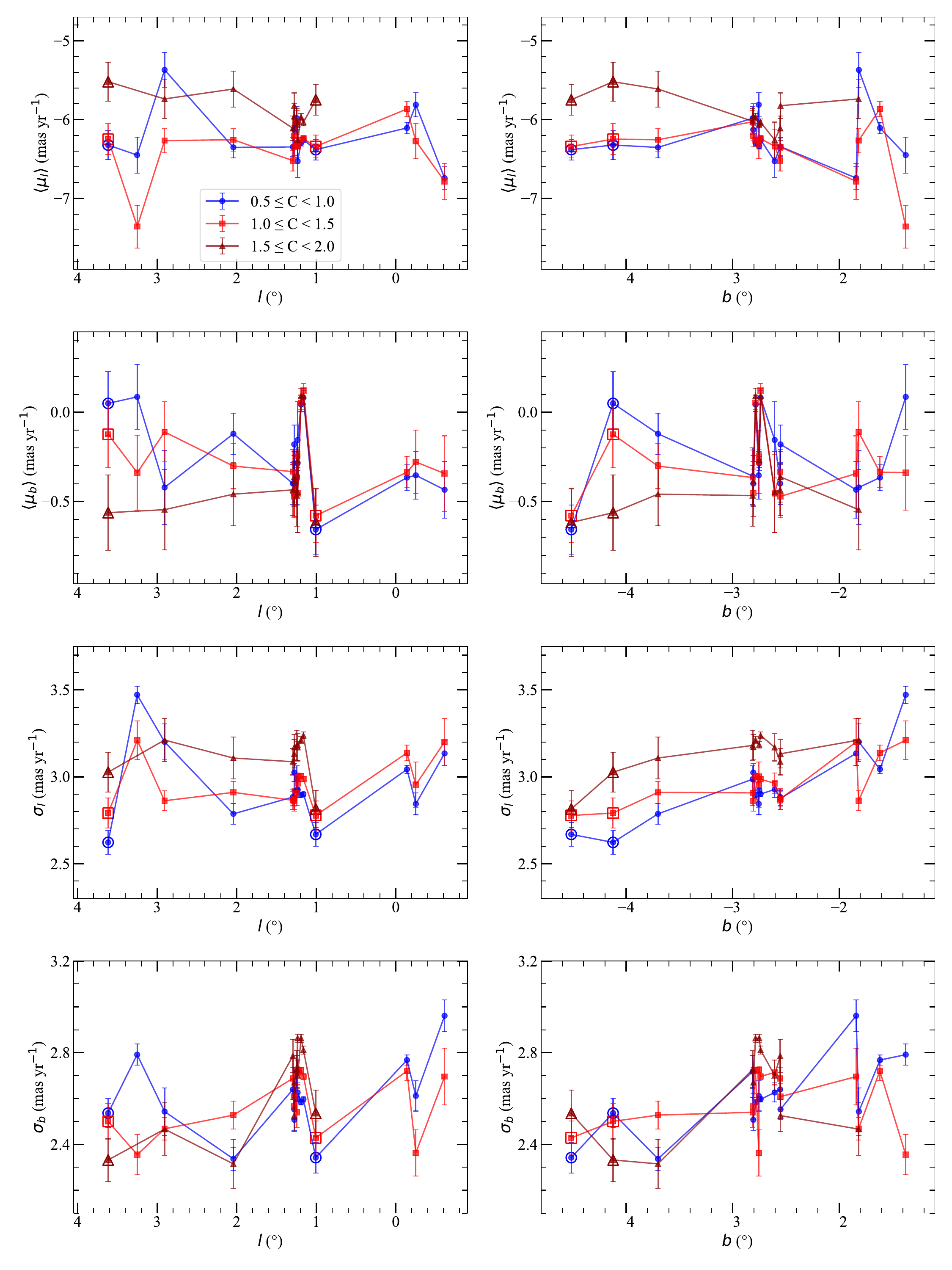}
  \caption{Reddening-corrected colour $C'_{\text{F606W}-\text{F814W}}$ (defined at the end of Section~\ref{sec:extinction_reddening}) ranges: $0.5 \le C'_{\text{F606W}-\text{F814W}} < 1.0$ (blue circles), $1.0 \le C'_{\text{F606W}-\text{F814W}} < 1.5$ (red squares), $1.5 \le C'_{\text{F606W}-\text{F814W}} < 2.0$ (dark red triangles). Fields 26 and 27 (marked with borders) are not corrected for reddening due to the small number of bright stars. Data points with fewer than 100 stars are not included.}
  \label{fig:pm_color}
\end{figure*}

\subsection{Non-Gaussianity of proper motion distributions}\label{sec:NonGaussianity}

Typically, the proper motion distribution of a stellar population can be described by a single Gaussian. However, we find that in some fields the distribution does not follow a single Gaussian but requires a superposition of two Gaussians. This may be due to the kinematic differences between bulge stars and foreground disc stars. In Section~\ref{sec:CMD}, two branches appear above the turn-off point in the CMD, with the blue bright branch thought to be mainly disc stars. To test whether the non-Gaussianity originates from disc contamination, we compare the proper motion distributions of this disc branch with those of the remaining stars.

Fig.~\ref{fig:pm_distribution_27} shows the absolute proper motion distributions of disc stars and the remaining (bulge) stars in Field 14. Both $\mu_l$ and $\mu_b$ clearly show non-Gaussian features. We fit each distribution with a double-Gaussian function. For the disc stars, the two Gaussian mean values for $\mu_l$ are $-6.410 \pm 2.566$ and $-1.641 \pm 0.706$~mas yr$^{-1}$, and for $\mu_b$ they are $-0.744 \pm 0.035$ and $-0.415 \pm 0.046$~mas yr$^{-1}$. For the bulge stars, the means for $\mu_l$ are $-6.746 \pm 0.019$ and $-1.623 \pm 0.053$~mas yr$^{-1}$, and for $\mu_b$ they are $0.050 \pm 0.006$ and $0.107 \pm 0.013$~mas yr$^{-1}$. The means of the two Gaussians are similar between disc and bulge samples, indicating that both samples consist of two subpopulations with different kinematic properties. One subpopulation has mean proper motion close to that of Sgr A* (i.e., bulge stars), while the other has a mean $\mu_l$ shifted positively away from Sgr A* (i.e., disc stars). Since disc motion is mainly along the Galactic plane, $\mu_l$ is more affected by disc contamination, while $\mu_b$ is less affected, which explains why the two Gaussian mean values differ more in $\mu_l$ than in $\mu_b$. Moreover, the relative fractions of the two subpopulations differ between disc and bulge samples: the disc sample has a larger fraction of the component whose mean proper motion deviates from Sgr A*, while the bulge sample has a larger fraction of the component close to Sgr A*. This further supports that the non-Gaussianity originates from disc contamination.

\citet{Clarkson2008BulgeProperMotions} analysed the proper motion distributions of stars near the Sagittarius Window ($l = 1.25^{\circ}$, $b = -2.65^{\circ}$) using HST SWEEPS data and found non-Gaussian features in disc branch stars. They found a difference of 4.17~mas yr$^{-1}$ between the two populations. In Baade's Window, \citet{Marchetti2022BDBSCleaning} also found a double Gaussian structure, and fitted bulge and residual foreground components with mean $\mu_l$ values of $-6.525$ and $-1.781$~mas yr$^{-1}$, respectively. These results also support that disc contamination is the dominant cause of the non-Gaussianity.

\subsection{Kinematics of stars in different magnitude and colour ranges}

To further investigate the dependence of bulge stellar kinematics on observed stellar properties, we divide the bulge stars into groups by magnitude and colour, and analyse the spatial distributions of mean proper motions and dispersions for each group. Before grouping, we remove the bright disc stars in the upper-left CMD (Section~\ref{sec:CMD}), and then apply extinction and reddening corrections using the RC parameters derived from CMD fits (Section~\ref{sec:extinction_reddening}). Therefore, the data in this section are limited to the 16 fields with colour information. Fields 26 and 27 are not corrected because they have too few bright stars for reliable RC fits in their CMDs.

Fig.~\ref{fig:pm_mag} shows the spatial distributions of mean proper motions and dispersions for bulge stars in different extinction-corrected magnitude bins, and Fig.~\ref{fig:pm_color} shows the corresponding distributions for different reddening-corrected colour bins. In the $l$ direction, the mean proper motion tends to be less negative for brighter stars; the dispersion in $l$ tends to be larger for brighter stars and also larger for redder stars. No clear trends are seen in the $b$ direction with magnitude or colour. Similarly, the red-clump, red giant branch bump and asymptotic giant
branch bump studied by \citet{Clarke2019VIRAC3D} and the N-body simulation by \citet{Steven2022BulgePMRotationCurves} both gave more positive mean $\mu_l$ at closer distances and brighter magnitudes across all latitudes.

% !TEX root = ../mnras_template.tex

\section{Conclusions}\label{chap:05conclusion}

In this work, we use the long-baseline archival observations from the WFC3/UVIS camera on HST, combined with data from \textit{Gaia} DR3, to perform high-precision absolute proper motion measurements and detailed statistical analyses of stars in the southern Galactic bulge. We select 27 fields in the southern bulge with repeated multi-epoch observations spanning 3 to 13 years, and measure the absolute proper motions of about 400,000 stars, achieving a precision of $\sim0.1$~mas yr$^{-1}$ for relatively bright stars. Using red clump stars as indicators, we perform colour–magnitude diagram (CMD) analyses of the bulge populations, effectively correcting for extinction and reddening in a subset of fields and removing contamination from foreground disc stars. We then carry out a detailed analysis of the proper motion distributions, including mean proper motions, dispersions, anisotropy, and covariance. Thanks to the high sensitivity of WFC3/UVIS, which enables proper motion measurements for a large number of faint stars ($18.5 < m_{\text{F814W}} < 23$), we measure these statistical quantities with significantly smaller errors.

We compare the proper motions of bulge stars with those of the supermassive black hole Sgr A* at the Galactic centre. In the $b$ direction, the mean proper motion of stars is in excellent agreement with that of Sgr A*. In the $l$ direction, however, the mean proper motion deviates from that of Sgr A* due to contamination by foreground disc stars, which move predominantly along the Galactic plane. We also find that the mean proper motion in $l$ shows a clear trend towards the Sgr A* value as we approach the Galactic plane. By extrapolating the linear fit to $b=0^{\circ}$, we obtain $\langle\mu_l\rangle = -6.383 \pm 0.065$~mas yr$^{-1}$, which is consistent with the Sgr A* measurement of $-6.379 \pm 0.026$~mas yr$^{-1}$ from \citet{Reid2004SgrAProperMotion}. This indicates that the extrapolated trend reflects the true spatial variation of the proper motions, and also confirms that the proper motion of Sgr A* can effectively represent the motion of the bulge centre.

The proper motion dispersions show even clearer spatial gradients. Both $\sigma_l$ and $\sigma_b$ exhibit significant increases towards the centre, and display a clear asymmetry between the two sides of the bulge ($l>0^{\circ}$ vs. $l<0^{\circ}$). The ratios of the slopes of the linear fits on the two sides are significantly less than 1, indicating that the gradient of the dispersion on the $l<0^{\circ}$ side is steeper. This dynamically implies the presence of non-axisymmetric structures, such as a bar, in the bulge. Further analysis reveals significant velocity anisotropy and spatial variations in the covariance of the bulge stars. The anisotropy parameter $\sigma_l/\sigma_b$ has a mean value of $1.131 \pm 0.005$, with all fields showing values greater than 1, indicating that the random motion is stronger in the $l$ direction (parallel to the Galactic plane). The normalised covariance $C_{lb}$ is negative on the $l>0^{\circ}$ side (mean $\sim -0.085$), while on the $l<0^{\circ}$ side it is essentially zero (mean $\sim 0.004$) with local positive values. The covariance and its spatial trends also point to non-axisymmetric features induced by the bar. Comparing our results with those of \citet{Kozlowski2006kinematics}, who measured bulge proper motions using HST/WFPC2 data, we find general agreement but some differences in the fitted trends. We attribute these differences primarily to the different sample selections and data qualities: the \citet{Kozlowski2006kinematics} sample was smaller and concentrated on brighter stars, whereas our larger and more precise data set allows us to better reveal the internal dynamical structure of the bulge.

Analysis of individual fields shows that the absolute proper motion distributions in some fields are clearly non-Gaussian and can be well described by a superposition of two Gaussians. By identifying the blue bright branch in the CMD, which is mainly composed of disc stars, we demonstrate that the non-Gaussianity mainly arises from foreground disc contamination. Double-Gaussian fitting of the proper motion components recovers the kinematic differences between the bulge and disc subpopulations. The disc peak in $\mu_l$ deviates from the Sgr A* value by about 4~mas yr$^{-1}$. This is consistent with the findings of \citet{Clarkson2008BulgeProperMotions}, further confirming that disc contamination is the primary source of the observed non-Gaussianity. After removing the blue bright branch and applying extinction and reddening corrections using the RC stars in the CMD, we examine the proper motion distributions in different magnitude and colour bins. We find that the mean proper motion and dispersion in the $l$ direction show some correlation with stellar brightness and colour, while those in the $b$ direction do not show any clear trends.

In summary, this work takes advantage of the high spatial resolution and long time baselines of HST, combined with the absolute reference frame provided by \textit{Gaia} DR3, to overcome the limitations of previous deep-field studies that were either restricted to relative measurements or had limited sample sizes. We obtain high-precision kinematic parameters for the bulge region. Future work may be desirable to compare our results with theoretical models of the Galactic bulge, such as the Besançon Galaxy model provided by \citet{Robin2012BulgePop2MASS}, and the made-to-measure models from \citet{Portail2017MadeToMeasure} and \citet{Clarke2022PatternSpeed}. This will help to further constrain the internal dynamical structure and formation history of the bulge.

% \clearpage
\section*{Acknowledgements}

Z.W., S.M., H.Y. and W.Z. acknowledge support by the National Natural Science Foundation of China (Grant No. 12133005). We thank Sean K. Terry for his help in HST astrometry. The authors thank the High Performance Computing Center at Westlake University for providing computational and data-storage resources.

This research is based on observations made with the NASA/ESA Hubble Space Telescope and retrieved from the Mikulski Archive for Space Telescopes (MAST) at the Space Telescope Science Institute (STScI). STScI is operated by the Association of Universities for Research in Astronomy, Inc., under NASA contract NAS5-26555. These observations are associated with HST programmes 11664, 11707, 12322, 12541, 12586, 13417, 13458, 13463, 13758, 15318, 15455, 15513, 16067, 16200, 16716, 16760, 16782, 17834, and 17838.

This work has made use of data from the European Space Agency (ESA) mission \textit{Gaia} (\url{https://www.cosmos.esa.int/gaia}), processed by the \textit{Gaia} Data Processing and Analysis Consortium (DPAC, \url{https://www.cosmos.esa.int/web/gaia/dpac/consortium}). Funding for the DPAC has been provided by national institutions, in particular the institutions participating in the \textit{Gaia} Multilateral Agreement.

The OGLE project has received funding from the Polish National Science Centre grant OPUS 2024/55/B/ST9/00447 awarded to A.U.

The authors acknowledge the use of ChatGPT (OpenAI) to assist with text refinement. The authors take full responsibility for the content.

%%%%%%%%%%%%%%%%%%%%%%%%%%%%%%%%%%%%%%%%%%%%%%%%%%
\section*{Data Availability}

The HST/WFC3 observations used in this work are publicly available from MAST under the programme IDs listed in the Acknowledgements. The Gaia DR3 data are publicly available from the Gaia Archive. The HST-derived stellar proper motion catalogue underlying the main results will be shared on reasonable request to the corresponding author. The OGLE/Uranus proper motion catalogue is not publicly available.



\bibliographystyle{mnras}
\bibliography{references}

%\begin{thebibliography}{99}
%\end{thebibliography}

%%%%%%%%%%%%%%%%%%%%%%%%%%%%%%%%%%%%%%%%%%%%%%%%%%

%%%%%%%%%%%%%%%%% APPENDICES %%%%%%%%%%%%%%%%%%%%%

% \appendix

% \section{Some extra material}

% If you want to present additional material which would interrupt the flow of the main paper,
% it can be placed in an Appendix which appears after the list of references.

%%%%%%%%%%%%%%%%%%%%%%%%%%%%%%%%%%%%%%%%%%%%%%%%%%

% Don't change these lines
\bsp	% typesetting comment
\label{lastpage}
\end{document}